\documentclass[12pt]{article}

\usepackage{amsmath,amssymb,amsthm,mathtools,bm}
\usepackage{hyperref}
\theoremstyle{plain}

\theoremstyle{definition}

\theoremstyle{proposition}

\theoremstyle{lemma}

\theoremstyle{remark}

\usepackage[pdftex]{graphicx}

\makeatletter
 
  \@addtoreset{equation}{section}
 \makeatother

\newcommand{\pd}{\partial}

\newcommand{\bea}{\begin{eqnarray}}
\newcommand{\eea}{\end{eqnarray}}

\def\X5sp{{\rm X}_5}
\def\Y3sp{{\rm Y}_3}
\def\Z3sp{{\rm Z}_3}

\def\mb#1{\mbox{\boldmath $#1$}} 

\begin{document}
\setlength{\oddsidemargin}{0cm}
\setlength{\baselineskip}{7mm}

\begin{titlepage}
\begin{flushright}   \end{flushright} 

~~\\

\vspace*{0cm}
    \begin{Large}
       \begin{center}
         {The heterotic  perturbative  vacua
         in string geometry theory}
       \end{center}
    \end{Large}
\vspace{1cm}

\begin{center}
	Koichi N{\sc agasaki},$^{*}$\footnote
           {e-mail address : koichi.nagasaki24@hirosaki-u.ac.jp} and 
        Matsuo S{\sc ato}$^{*}$\footnote
           {e-mail address : msato@hirosaki-u.ac.jp}
\\
      \vspace{1cm}
       
         {$^{*}$\it Graduate School of Science and Technology, Hirosaki University\\ 
 Bunkyo-cho 3, Hirosaki, Aomori 036-8561, Japan}\\

%
                    
\end{center}

\hspace{5cm}

\begin{abstract}
\noindent
String geometry theory is one of the candidates of the non-perturbative formulation of superstring theory. In this paper, in string geometry theory, we identify perturbative  heterotic vacua, which include general  heterotic backgrounds. From fluctuations around these vacua, we derive the path-integrals of heterotic perturbative superstrings on the backgrounds up to any order. 

\end{abstract}

\vfill
\end{titlepage}
\vfil\eject

\setcounter{footnote}{0}

\section{Introduction}\label{intro}
\setcounter{equation}{0}

String geometry theory is defined by a path-integral of string manifolds, which are a class of infinite-dimensional manifolds \cite{Sato:2017qhj}, and are defined by patching open sets of the model space defined by introducing a topology to a set of superstrings.
Although the theory is defined by a path-integral of string manifolds, there is no problem of non-renormalizability, 
because a non-renormalization theorem in string geometry theory states that there is no ``loop'' correction \cite{Sato:2025wfc}, controlled by ``quantum'' correction parameter $\beta$ in the path-integral of string geometry theory, which is independent of quantum correction parameter $\hbar$ in string theory. We distinguish the effects of $\beta$ and $\hbar$ by putting " " like "classical" and "loops" for tree level and loop corrections with respect to $\beta$, respectively, whereas by putting nothing like classical and loops for tree level and loop corrections with respect to $\hbar$, respectively. 
From the string geometry theory in the ``tree'' level, 
the path-integrals of  perturbative superstrings are derived up to any order in $\hbar$, including the moduli of super Riemann surfaces \cite{Sato:2017qhj, Sato:2019cno,  Sato:2020szq}, because string geometry includes information of genera of the world-sheets of the strings.


So far, we derived the path-integrals of perturbative superstrings on the superstring backgrounds that consist of the flat background and the first order expansions around it, from string geometry theory \cite{Sato:2022owj, Sato:2022brv}.  In this case, we can identify superstring backgrounds up to  only the first order and an effective potential for the backgrounds becomes trivial. In this paper, we will derive  the path-integrals of heterotic perturbative superstrings on  heterotic superstring backgrounds that consist of the flat background and the all  order expansions around it. As a result, we identify the heterotic perturbative vacua completely.


This paper is a supersymmetric generalization of \cite{Nagasaki:2023fnz}. The bosonic theory cannot be used to investigate our four-dimensional physics because of its tachyonic states, although it is suitable for the first step to study the full theory because it is rather easy to study.  
On the other hand, heterotic vacua are phenomenologically very important. The discovery of the heterotic string in \cite{Gross:1984dd} enabled string phenomenology through the ground unification theory, where people found many heterotic string phenomenological models that may include the Standard Model of elementary physics although they are not complete.

The organization of the paper is as follows. 
In section 2, we briefly review the heterotic sector of string geometry theory. 
In section 3,  we set heterotic perturbative superstring vacua  parametrized by the heterotic  superstring backgrounds $G_{\mu\nu}(x)$, $B_{\mu\nu}(x)$, $\phi(x)$, and $A_{\mu}(x)$, and consider fluctuations around the vacua. As a result, we derive the path-integrals of  heterotic perturbative superstrings on the  backgrounds.
In this process, we derive conditions to determine heterotic perturbative superstring vacua. In section 4,  we obtain a potential for heterotic superstring backgrounds by substituting the heterotic perturbative superstring vacua into the ``classical'' potential in string geometry theory and imposing the conditions  by the method of Lagrange multipliers. For applications to string phenomenology, we further obtain a potential on a restricted region of string backgrounds with the warped compactification. In section \ref{sec:discussion}, we conclude and discuss our results.

\vspace{1cm}
\section{Brief review on the heterotic sector of string geometry theory}\label{sec:rev_stringgeometry}
String geometry theory is a natural non-pertubative generalization of the perturbative sting theory, where particles consist of strings. Furthermore, the space-time is also consist of strings in string geometry theory.  The motivation for this is given as follows. It has not succeeded to obtain ordinary relativistic quantum gravity that is defined by a path integral over metrics on a space representing the spacetime itself because of ultraviolet divergences. The reason would be impossibility to regard points as fundamental constituents of the spacetime because the spacetime itself fluctuates at the Plank scale.  Thus, it is reasonable to define quantum gravity by a path integral over metrics on a space that consists of strings, by making a point have a structure of strings. In fact, perturbative strings are shown to suppress the ultraviolet divergences in quantum gravity.

In string geometry theory, we geometrically define a space of superstrings including the effect of interactions. For this purpose, here we first review how such spaces of strings are defined in string field theories. In these theories, after a free loop space of strings are prepared, interaction terms of strings in actions are defined. In other words, the spaces of strings are defined by deforming the ring on the free loop space. Geometrically, the space of strings is defined by deformation quantization of the free loop space as a noncommutative geometry. Actually, in Witten's cubic open string field theory \cite{WittenCubic}, the interaction term is defined by using the $*$-product of noncommutative geometry. 
On the other hand, we adopt different approach, namely (infinite-dimensional) manifold theory\footnote{See \cite{RiemannianGeometry} as an example of text books for infinite-dimensional manifolds.}. We do not start with a free loop space, but we define a space of strings including the effect of interactions from the beginning. This is realized by defining the space of strings as a collection of world-time constant lines of Riemann surfaces.  
The criterion to define a topology, which represents how near the strings are, is that trajectories in asymptotic processes on the space of strings reproduce the right moduli space of the Riemann surfaces in a target manifold. We need Riemannian geometry naturally for fields on the space of strings because it is not flat.

String manifold is constructed by patching open sets in string model space $E= \cup_{T} E_T$, where $T$ runs  IIA, IIB, SO(32) I, SO(32) het, and $E_8 \times E_8$ het.
Here, we summarize the definition of the heterotic sector of the string model space, $E_{G \mbox{het}}$ where $G$ runs $SO(32)$ and $E_8 \times E_8$. 
First, one of the coordinates of the model space is spanned by string geometry time $\bar{\tau} \in \bold{R}$ and another is spanned by heterotic super Riemann surfaces $\bar{\bold{\Sigma}} \in \mathcal{M}_{\mbox{het}}$ \cite{NotesOnSupermanifolds, WittenSupermoduli, SuperPeriod}. 
On each  $\bar{\bold{\Sigma}}$,  
 a global time is defined canonically and uniquely by the real part of the integral of an Abelian differential \cite{Krichever:1987a, Krichever:1987b}.
We identify this global time as $\bar{\tau}$ and restrict $\bar{\bold{\Sigma}}$ to a $\bar{\tau}$ constant hyper surface, and obtain $\bar{\bold{\Sigma}}|_{\bar{\tau}}$. 
An embedding of $\bar{\bold{\Sigma}}|_{\bar{\tau}}$ to $\mathbb{R}^{d}$  is parametrized by the other coordinates 
${\boldsymbol X}_{G}^{(\mu \bar\sigma  \bar\theta)}(\bar{\tau})=X^{\mu}(\bar\sigma, \bar\tau)+ \bar{\theta} \psi^{\mu}(\bar\sigma, \bar\tau) $ where $\mu=0, 1, \cdots d-1$ and $\psi^{\mu}$ is a Majorana fermion, and $\bm X_{LG}^{(A \bar\sigma  \bar\theta^-)}(\bar\tau) =  \bar\theta^{-} \lambda_{G}^A(\bar\sigma, \bar\tau)$ where $A= 1, \cdots 32$ and 
$\bar\theta^{-}$ has the opposite chirality to $\bar\theta$. 
We abbreviate $G$ of $X^{\mu}$ and $\psi^{\mu}$. 

We can define worldsheet fermion numbers of states in a Hilbert space because the states consist of the fields over the local coordinates ${\boldsymbol X}_{G}^{(\mu \bar\sigma  \bar\theta)}(\bar{\tau})=X^{\mu}(\bar\sigma, \bar\tau)+ \bar{\theta} \psi^{\mu}(\bar\sigma, \bar\tau)$ and $\bm X_{LG}^{(A \bar\sigma  \bar\theta^-)}(\bar\tau) =  \bar\theta^{-} \lambda_{G}^A(\bar\sigma, \bar\tau)$. 
For $G=SO(32)$, we take periodicities 
\begin{equation}
\lambda_{SO(32)}^A(\bar{\tau}, \bar{\sigma}+2\pi)=\pm \lambda_{SO(32)}^A(\bar{\tau}, \bar{\sigma}) \quad (A=1, \cdots 32)
\end{equation}
with the same sign on all 32 components. 
We define the Hilbert space in these coordinates by the GSO projection of the states with $e^{\pi i F}=1$ and $e^{\pi i \tilde{F}}=1$, where $F$ and $\tilde{F}$ are the numbers of left- and right- handed fermions $\lambda^A_{SO(32)}$ and $\psi^{\mu}$, respectively. 
For $G=$ $E_8 \times E_8$, the periodicity is given by 
\begin{eqnarray}
\lambda_{E_8 \times E_8}^A(\bar{\tau}, \bar{\sigma}+2\pi)= 
\left\{
\begin{array}{c}
\eta \lambda_{E_8 \times E_8}^A(\bar{\tau}, \bar{\sigma}) \quad (1 \leqq A \leqq 16) \\
\eta' \lambda_{E_8 \times E_8}^A(\bar{\tau}, \bar{\sigma}) \quad (17 \leqq A \leqq 32), 
\end{array}
\right.
\end{eqnarray}
with the same sign $\eta(= \pm 1)$ and $\eta'(= \pm 1)$ on each 16 components.  
The GSO projection is given by $e^{\pi i F_1}=1$, $e^{\pi i F_2}=1$ and $e^{\pi i \tilde{F}}=1$, where $F_1$, $F_2$ and $\tilde{F}$ are the numbers of $\lambda_{E_8 \times E_8}^{A_1}$ ($A_1=1, \cdots, 16$), $\lambda_{E_8 \times E_8}^{A_2}$ ($A_2=17, \cdots, 32$) and $\psi^{\mu}$, respectively.

\begin{figure}[htb]
\centering
\includegraphics[width=3cm]{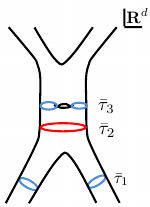}
\caption{Various string states. The red and blue lines represent one string and two strings, respectively.}
\label{states}
\end{figure}

Because the bosonic part of $\bar{\bold{\Sigma}}|_{\bar{\tau}}$ is isomorphic to $ S^1 \cup S^1 \cup \cdots \cup S^1$ and $\bold{X}_{G}(\bar{\tau}): \bar{\bold{\Sigma}}|_{\bar{\tau}} \to \mathbb{R}^{d}$, $[\bar{\bold{\Sigma}},  \bold{X}_{G}(\bar{\tau}), \lambda_{G}(\bar{\tau}), \bar{\tau}]$  represent many-body strings in $\mathbb{R}^{d}$ as in Fig. \ref{states}.
Heterotic model space $E_{G \mbox{het}}$  is defined by the collection of $[\bar{\bold{\Sigma}},  \bold{X}_{G}(\bar{\tau}), \lambda_{G}(\bar{\tau}), \bar{\tau}]$ by considering all the  $\bar{\bm\Sigma}$, all the values of $\bar{\tau}$, and all the $\bold{X}_{G}(\bar{\tau})$ and $\lambda_{G}(\bar{\tau})$: 
$E_{G \mbox{het}} = \{[\bar{\bold{\Sigma}},  \bold{X}_{G}(\bar{\tau}), \lambda_{G}(\bar{\tau}), \bar{\tau}]\}$.

How near the two string states is defined by how near the values of $\bar{\tau}$, $\bm{X}_{\hat{D}_{G}}(\bar{\tau})$ and  $\lambda_{G}(\bar{\tau})$. An $\epsilon$-open neighborhood of 
$[\bar{\bold{\Sigma}},  \bold{X}_{s G}(\bar{\tau}_s), \lambda_{s G}(\bar{\tau}_s), \bar{\tau}_s]$ is given by
\begin{eqnarray}
&&U([\bar{\bold{\Sigma}},  \bold{X}_{s G}(\bar{\tau}_s), \lambda_{s G}(\bar{\tau}_s), \bar{\tau}_s], \epsilon) \nonumber \\
&:=&
\biggl\{[\bar{\bold{\Sigma}},  \bold{X}_{G}(\bar{\tau}), \lambda_{G}(\bar{\tau}), \bar{\tau}]
\bigm| \nonumber \\
&&\sqrt{|\bar{\tau}-\bar{\tau}_s|^2
+\| \bold{X}_{G}(\bar{\tau}) -\bold{X}_{s G}(\bar{\tau}_s) \|^2
+\| \lambda_{G}(\bar{\tau})
- \lambda_{s G}(\bar{\tau}_s)\|^2}
<\epsilon  \biggr\}, \label{HeteroNeighbour}
\end{eqnarray}
where 
\begin{eqnarray}
&&
\| \bold{X}_{G}(\bar{\tau}) -\bold{X}_{s G}(\bar{\tau}_s) \|^2 \nonumber \\
&:=&\int_0^{2\pi}  d\bar{\sigma} 
\Bigl(|x(\bar{\tau}, \bar{\sigma})-x_s(\bar{\tau}_s, \bar{\sigma})|^2 
+(\bar{\psi}(\bar{\tau}, \bar{\sigma})-\bar{\psi}_s(\bar{\tau}_s, \bar{\sigma}))
(\psi(\bar{\tau}, \bar{\sigma})-\psi_s(\bar{\tau}_s, \bar{\sigma})) \Big)\nonumber \\
&&\| \lambda_{G}(\bar{\tau})
- \lambda_{s G}(\bar{\tau}_s)\|^2 
\nonumber \\
&:=&\int_0^{2\pi}  d\bar{\sigma} 
( \bar{\lambda}_{G}(\bar{\tau}, \bar{\sigma})-\bar{\lambda}_{s G}(\bar{\tau}_s, \bar{\sigma}))
(\lambda_{G}(\bar{\tau}, \bar{\sigma})-\lambda_{s G}(\bar{\tau}_s, \bar{\sigma})),
\end{eqnarray}
where $\bar{\bold{\Sigma}}$  is a discrete variable in the topology of string geometry.
In this neighborhood,  $\bar{\tau}$, $\bold{X}_{G}(\bar{\tau})$ and $\lambda_{G}(\bar{\tau})$ have the same weights because we impose diffeomorphism invariance that mixes $\bar{\tau}$, $\bold{X}_{G}(\bar{\tau})$ and $\lambda_{G}(\bar{\tau})$  completely to the theory so that it has the maximal symmetry.
The precise definition of the string topology is given in  \cite{Sato:2017qhj, Sato:2025wfc}. 
By this definition, arbitrary two string states on a connected super Riemann surface in $E$ are connected continuously. 
Thus, there is a one-to-one correspondence between a super Riemann surface in $\mathbb{R}^{d}$ and a curve  parametrized by $\bar{\tau}$ from $\bar{\tau}=-\infty$ to $\bar{\tau} = \infty$ on $E$. 
That is, curves that represent asymptotic processes on $E_{G \mbox{het}}$ reproduce the right moduli space of the heterotic super Riemann surfaces in $\mathbb{R}^{d}$. 
Therefore, a string geometry theory possesses all-order information of superstring theory. 
Indeed, the path integral of perturbative superstrings is derived from the string geometry theory as in \cite{Sato:2017qhj, Sato:2019cno, Sato:2020szq, Sato:2022owj, Sato:2022brv, Nagasaki:2023fnz}. 
The consistency of the perturbation theory determines $d=10$ (the critical dimension).

In the following,  instead of $[\bar{{\boldsymbol \Sigma}}, \bold{X}_{G}(\bar{\tau}), \lambda_{G}(\bar{\tau}), \bar{\tau}]$,  we denote $[\bar{\bold{E}}_{M}^{\quad A}(\bar{\sigma}, \bar{\tau}, \bar{\theta}), \bold{X}_{G}(\bar{\tau}), \lambda_{G}(\bar{\tau}), \bar{\tau}]$, 
where $\bar{\bold{E}}_{M}^{\quad A}(\bar{\sigma}, \bar{\tau}, \bar{\theta}^+)$ ($M=(m, +)$, $A=(q, +)$, $m, q=0,1$, $\theta^+ :=\theta$) is the worldsheet super vierbein on $\bar{{\boldsymbol \Sigma}}$ \cite{Brooks:1986uh}, 
because giving a super Riemann surface is equivalent to giving a super vierbein up to super diffeomorphism and super Weyl transformations.

The summations over $(\bar\sigma, \bar\theta)$ and $(\bar\sigma',\bar\theta^{-})$ are defined by 
$\displaystyle\int d\bar\sigma d\bar\theta\hat{\bm E}(\bar\sigma, \bar\tau, \bar\theta)$ and 
$\displaystyle\int d\bar\sigma' d\bar\theta^{-}\bar e(\bar\sigma', \bar\tau)$, respectively. $\hat{\bm E}(\bar{\sigma}, \bar{\tau}, \bar{\theta})\coloneqq 
(1/\bar{n})\bar{\bm E}(\bar{\sigma}, \bar{\tau}, \bar{\theta})$, where $\bar n$ is the lapse function of the two-dimensional metric (See \eqref{eq:def_ADHM_matrix}).
These summations are transformed as scalars under $\bar\tau \mapsto \tilde{\bar{\tau}}(\bar\tau, \bm X_{G}(\bar\tau), \bm X_{L G}(\bar\tau))$.  Moreover, $\displaystyle\int d\bar\sigma d\bar\theta\hat{\bm E}(\bar\sigma, \bar\tau, \bar\theta)$ is invariant under a supersymmetry transformation $(\bar{\sigma}, \bar{\theta}) \mapsto (\tilde{\bar{\sigma}}(\bar{\sigma}, \bar{\theta}), \tilde{\bar{\theta}}(\bar{\sigma}, \bar{\theta}))$.
$\displaystyle\int d\bar\sigma' d\bar\theta^{-}\bar e(\bar\sigma', \bar\tau)$ is also invariant under this supersymmetry transformation, because 
$(\mu\bar\sigma\bar\theta)$ and $(A\bar\sigma' \bar\theta^{-})$ in $I = \{d,(\mu\bar\sigma\bar\theta),(A\bar\sigma' \bar\theta^{-})\}$
are independent indices and then $(A\bar\sigma' \bar\theta^{-})$ is not transformed under the supersymmetry.  
As a result, the heterotic part of any action is invariant under this $\mathcal N = (1,0)$ supersymmetry transformation because all the indices are contracted by the summations.

The cotangent space is spanned by
\begin{align}
d\bm X_{G}^d &\coloneqq d\bar\tau,\nonumber\\
d\bm X_{G}^{(\mu\bar\sigma\bar\theta)} 
 &\coloneqq d\bm X_{G}^\mu(\bar\sigma,\bar\tau,\bar\theta),\nonumber\\
d\bm X_{LG}^{(A\bar\sigma\bar\theta^{-})} 
 &\coloneqq d\bm X_{LG}^A(\bar\sigma,\bar\tau,\bar\theta^-). \label{cotangent}
\end{align}
 $d \bar{\bm E}$   cannot be a part of basis that span the cotangent space because $ \bar{\bm E}$ is a discrete variable as in (\ref{HeteroNeighbour}). 
An explicit form of the line element is given in the same way as in the finite dimensional case by
\begin{align}\label{LineElement}
&ds^2(\bar{\bm E}, \bm X_{G}(\bar\tau), \bm X_{LG}(\bar\tau), \bar\tau)\nonumber\\
&= G(\bar{\bm E}, \bm X_{G}(\bar\tau),  \bm X_{LG}(\bar\tau), \bar\tau)_{dd} (d\bar\tau)^2\nonumber\\ 
&\quad
+ 2d\bar\tau\int d\bar\sigma d\bar\theta\hat{\bm E}
\sum_\mu G(\bar{\bm E}, \bm X_{G}(\bar\tau), \bm X_{LG}(\bar\tau), \bar\tau)_{d(\mu \bar\sigma\bar\theta)} d\bm X_{G}^{\mu}(\bar{\sigma}, \bar{\tau}, \bar{\theta}) \nonumber \\ 
&\quad
 +2 d\bar\tau\int d\bar\sigma d\bar\theta^{-} \bar e\sum_AG(\bar{\bm E}, \bm X_{G}(\bar\tau), \bm X_{LG}(\bar\tau,\bar\tau)_{d(A\bar\sigma\bar\theta^{-})} d \bm X_{LG}^A(\bar\sigma, \bar\tau, \bar\theta^{-})\nonumber\\
&\quad
 +\int d\bar{\sigma} d\bar{\theta} \hat{\bm E}  \int d\bar\sigma'd\bar\theta'\hat{\bm E}'   
 \sum_{\mu, \mu'} G(\bar{\bm E},\bm X_{\hat D_{T}}(\bar\tau), \bm X_{LG}(\bar\tau), \bar{\tau})_{ \; (\mu \bar{\sigma} \bar{\theta})  \; (\mu' \bar{\sigma}' \bar{\theta}')} d\bm X_{G}^{\mu}(\bar{\sigma}, \bar{\tau}, \bar{\theta})d\bm X_{G}^{\mu'}(\bar\sigma',\bar\tau, \bar\theta')\nonumber\\
&\quad
 +\int d\bar\sigma d\bar{\theta}\hat{\bm E}\int d\bar\sigma' d\bar\theta^{-}\bar e'
  \sum_{\mu, A} G(\bar{\bm E}, \bm X_{G}(\bar\tau),\bm X_{LG}(\bar\tau), \bar\tau)_{(\mu \bar\sigma\bar\theta)(A\bar\sigma'\bar\theta^{-})}d\bm X_{G}^\mu(\bar\sigma,\bar\tau,\bar\theta) 
 d\bm X_{LG}^A(\bar\sigma',\bar\tau,\bar\theta^{-})\nonumber\\
&\quad
 +\int d\bar{\sigma} d \bar\theta^{-}\bar{e}  \int d\bar{\sigma}'  d \bar{\theta}^{'-} \bar{e}'  \sum_{A, A'} G(\bar{\bm E},\bm X_{G}(\bar\tau),\bm X_{LG}(\bar\tau),\bar\tau)_{(A \bar\sigma\bar\theta^{-})(A'\bar\sigma'\bar\theta'^{-})}
d\bm X_{LG}^A(\bar\sigma,\bar\tau,\bar\theta^{-})d\bm X_{LG}^{A'}(\bar\sigma', \bar\tau, \bar\theta'^{-}). 
\end{align}
Here, we should note that the fields are functionals of $\bar{\bm E}$. 
The inverse metric $\bm G^{IJ}(\bar{\bm E}, \bm X_{G}(\bar\tau), \bm X_{LG}(\bar\tau), \bar\tau)$ is given by 
$\bm G_{IJ}\bm G^{JK} = \bm G^{KJ}\bm G_{JI}= \delta^{K}_{I}$, where 
$\delta^{d}_{d}=1$,  
$\delta_{\mu \bar\sigma \bar\theta}^{\mu' \bar\sigma' \bar\theta'} 
 = \frac1{\hat{\bm E}}\delta_\mu^{\mu'}\delta(\bar\sigma - \bar\sigma') \delta(\bar\theta - \bar\theta')$, 
$\delta_{A \bar{\sigma} \bar{\theta}^-}^{A' \bar{\sigma}' \bar{\theta}^{'-}} 
 = \frac1{\bar e}\delta_{A}^{A'}\delta(\bar\sigma - \bar\sigma') \delta(\bar\theta^{-}-\bar\theta'^{-})$. 
The dimensions of string manifolds, which are infinite dimensional manifolds, are formally given by the trace of ``1'', $\delta^M_M = D+1$, where 
$\displaystyle D\coloneqq \int d\bar{\sigma} d\bar{\theta} \hat{\bm E} \delta^{(\mu\bar\sigma \bar\theta)}_{(\mu\bar\sigma\bar\theta)}+\int d\bar\sigma d\bar\theta^{-}\bar e \delta_{A \bar{\sigma} \bar{\theta}^-}^{A \bar{\sigma} \bar{\theta}^{-}}$. Thus, we treat $D$ as regularization parameter and will take $D \to\infty$ later. The scalar $\varPhi(\bar h, X(\bar\tau), \bar\tau)$
and tensors $\mathcal B_{IJ}(\bar h, X(\bar\tau), \bar\tau), \cdots$ are also defined in the same way as in the finite dimensional case because the basis of the cotangent space is given explicitly as (\ref{cotangent}).

String geometry theory is defined by a partition function
\begin{align}
Z = \int \mathcal D\bm G\mathcal D\bm\Phi  \mathcal{D}\bm B\mathcal D\bm A\mathcal D\bm{\mathcal C}e^{\frac{i}{\beta} S},
\label{pathint}
\end{align}
where the action is given by
\begin{equation}
S = \int\mathcal D\bar\tau \mathcal D\bm E\mathcal D\bm X_{T}\sqrt{-\bm{\mathcal G}} 
 \bigg(e^{-2\bm\varPhi} \Big(\bm{\mathcal R} 
 + 4\bm\nabla_I\bm\varPhi\bm\nabla^I\bm\varPhi 
 - \frac12\vert\tilde{\bm{\mathcal H}}\vert^{2} - \frac{\alpha'}{4}\mathrm{tr}(\vert\bm{\mathcal F}\vert^2)\Big)
 - \frac12 \sum_{p=1}^\infty \vert\tilde{\bm{\mathcal F}}_p\vert^2     
\bigg), 
\label{action of bos string-geometric model}
\end{equation}
where the parameter of ``quantum'' corrections $\beta$ in the path-integral of the theory is independent of that of quantum corrections $\hbar$ in the perturbative string theories.
We distinguish the effects of $\beta$ and $\hbar$ by putting " " like "classical" and "loops" for tree level and loop corrections with respect to $\beta$, respectively, whereas by putting nothing like classical and loops for tree level and loop corrections with respect to $\hbar$, respectively.
We use the Einstein notation for the index $I$ and
$|\tilde{\bm{\mathcal H}}|^2 =
 (1/3!)\bm{\mathcal G}^{I_1J_1}\bm{\mathcal G}^{I_2J_2}\bm{\mathcal G}^{I_3J_3} 
 \tilde{\bm{\mathcal H}}_{I_1I_2I_3} \tilde{\bm{\mathcal H}}_{J_1J_2J_3}$ for example.
The equations of motion of this action \eqref{action of bos string-geometric model} can be consistently truncated to the ones of all the ten-dimensional  supergravities, namely type IIA, IIB, SO(32) type I, and type  SO(32) and $E_8 \times E_8$ heterotic supergravities \cite{Honda:2020sbl, Honda:2021rcd}. That is, this model includes all the superstring backgrounds.  Moreover, the action \eqref{action of bos string-geometric model} is strongly constrained by T-symmetry in string geometry theory, which is a generalization of T-duality among perturbative vacua in superstring theory \cite{Sato:2023lls}.
The action consists of fields on Riemannian string manifolds: a scalar curvature  $\bm{\mathcal R}$ of a metric $\bm{\mathcal G}_{I_1I_2}$, a scalar field $\bm\varPhi$, $p$-forms $\tilde{\bm{\mathcal F}}_p = \bm{\mathcal F}_p+\bm{\mathcal H}_3\wedge\bm{\mathcal C}_{p-3}$ where  $\bm{\mathcal F}_p$ are field strengths of ($p-1$)-form fields $\bm{\mathcal C}_{p-1}$,  $\tilde{\bm{\mathcal H}} = \bm{\mathcal H} -\bm{\mathcal \omega}_3$ where $\bm{\mathcal H}$ is a field strength of  a two-form field $\bm{\mathcal B}$, 
$\bm{\mathcal \omega}_3 = \mathrm{tr}(\bm {\mathcal A}\wedge d\bm{\mathcal  A} 
 - (2i/3)\bm{\mathcal  A} \wedge\bm{\mathcal  A} \wedge\bm{\mathcal  A})$, and 
$\bm{\mathcal  A}$ is an $N \times N$ Hermitian gauge field, whose field strength is given by $\bm{\mathcal  F}$. It is natural that the backgrounds of perturbative string theory are included in the expectation values of the fields in a non-perturbative formulation of string theory.  Actually, the fundamental fields in string geometry theory are extensions of those in the ten-dimensional supergravities. In order to minimize the number of the fundamental fields, the theory does not include such extensions of the massive modes in string theory.   However, the massive modes are included non-trivially in the theory because the perturbative string theory is derived from string geometry theory as one can see in the next section.

In the following, we fix $\text{T} = SO(32) \text{ hetero or }E_8 \times E_8 \text{ hetero}$, namely we choose heterotic chatrs, where $\bm C_p=0$.
 Then, the action \eqref{action of bos string-geometric model} becomes 
\begin{equation}\label{eq:sec3_adction}
S = \int\mathcal D\bar\tau \mathcal D\bm E\mathcal D\bm X_{G}\mathcal D\bm X_{LG}
\sqrt{-\bm{\mathcal G}}e^{-2\bm\varPhi}
\Big({\bm{\mathcal R}} + 4\bm\nabla_I{\bm\varPhi}\bm\nabla^I{\bm\varPhi} 
 - \frac12 |\tilde{\bm{\mathcal H}}|^2 - \frac{\alpha'}{4}\mathrm{tr}|\bm{\mathcal F}|^2\Big).
\end{equation}
In these charts, 
$I = \{d,(\mu\bar\sigma\bar\theta),(A\bar\sigma\bar\theta^{-})\}$.

%
%
\section{Deriving the path-integrals of heterotic perturbative superstrings on superstring backgrounds}
In this section, we will derive the path integrals of heterotic perturbative superstrings up to any order from ``tree''-level two-point correlation functions of the scalar fluctuations of the metric in string geometry theory. 

We set ``classical'' backgrounds that  represent heterotic perturbative superstring vacua,
\begin{subequations}\label{eq:sec3_backgrounds}
\begin{align}
\bm{\mathcal G}_{IJ} &=\bar{\bm G}_{IJ}, 
\label{eq:sec3_backgrounds_G}\\
\bm{\mathcal B}_{IJ}& =\bar{\bm B}_{IJ} 
\label{eq:sec3_backgrounds_B}\\
\bm\varPhi &= \bar{\bm\varPhi}
\label{eq:sec3_backgrounds_Phi}\\
\bm{\mathcal A}_{I}& = \bar{\bm A}_{I},
\label{eq:sec3_backgrounds_C}
\end{align}
\end{subequations}
where
\begin{subequations}\label{eq:sec3_condition2}
\begin{align}
\bar{\bf G}_{dd}&= {\bf G}_{dd} = e^{2\phi[G,B,\Phi, A; X]}  \\
\bar{\bf G}_{d(\mu\bar{\sigma}\bar{\theta})}&=0 \\
\bar{\bf G}_{d(A\bar{\sigma}\bar{\theta}^-)}&=0 \\
\bar{\bf G}_{(\mu\bar{\sigma}\bar{\theta})(\mu'\bar{\sigma}'\bar{\theta}')}&=
{\bf G}_{(\mu\bar{\sigma}\bar{\theta})(\mu'\bar{\sigma}'\bar{\theta}')}
=
\frac{\bar{e}^3}{\sqrt{\bar{h}}}\,
G_{\mu\nu}({\bf X}_{G}(\bar{\sigma}, \bar{\theta}))
\delta_{\bar{\sigma}\bar{\sigma}'}\,\delta_{\bar{\theta}\bar{\theta}'} \\
\bar{\bf G}_{(\mu\bar{\sigma}\bar{\theta})(A\bar{\sigma}'\bar{\theta'}^-)}&=0\\
\bar{\bf G}_{(A\bar{\sigma}\bar{\theta}^-)(A'\bar{\sigma}'\bar{\theta'}^-)}&=
\bf G_{(A\bar{\sigma}\bar{\theta}^-)(A'\bar{\sigma}'\bar{\theta'}^-)}
=\frac{\bar{e}^3}{\sqrt{\bar{h}}}\,\delta_{AA'}\,
\delta_{\bar{\sigma}\bar{\sigma}'}\,\delta_{\bar{\theta}^-\bar{\theta'}^-}
\\
\bar{\bf B}_{d(\mu\bar{\sigma}\bar{\theta})}&=0  \\
\bar{\bf B}_{d(A\bar{\sigma}\bar{\theta}^-)}&=0 \\
\bar{\bf B}_{(\mu\bar{\sigma}\bar{\theta})(\mu'\bar{\sigma}'\bar{\theta}')}&=
{\bf B}_{(\mu\bar{\sigma}\bar{\theta})(\mu'\bar{\sigma}'\bar{\theta}')}
=\frac{\bar{e}^3}{\sqrt{\bar{h}}}\,
B_{\mu\nu}({\bf X}_{G}(\bar{\sigma}, \bar{\theta}))
\delta_{\bar{\sigma}\bar{\sigma}'}\,\delta_{\bar{\theta}\bar{\theta}'} \\
\bar{\bf B}_{(\mu\bar{\sigma}\bar{\theta})(A\bar{\sigma}'\bar{\theta'}^-)}&=0\\\bar{\bf B}_{(A\bar{\sigma}\bar{\theta}^-)(A'\bar{\sigma}'\bar{\theta'}^-)}&=0
\\
\bar{\varPhi}&=\varPhi= \int d \bar{\sigma}\,d\bar{\theta}\,\hat{\bf E}\,\Phi({\bf X}_{G}(\bar{\sigma}, \bar{\theta})),\\
\bar{\bf A}_{d}&=0 \\
\bar{\bf A}_{(\mu\bar{\sigma}\bar{\theta})}&=
{\bf A}_{(\mu\bar{\sigma}\bar{\theta})}
=
\frac{\bar{e}^3}{\sqrt{\bar{h}}}\,
A_{\mu}({\bf X}_{G}(\bar{\sigma}, \bar{\theta}))\\
\bar{\bf A}_{(A\bar{\sigma}\bar{\theta}^-)}&=0, 
\end{align}
 \end{subequations}
 where $G_{\mu\nu}(x)$, $B_{\mu\nu}(x)$, $\Phi(x)$, and $A_{\mu}(x)$ represent heterotic superstring backgrounds in the ten dimensions, and $\phi$ will be determined later. 
  Actually, it was shown that an infinite number of equations of motion of string geometry theory are consistently truncated by these configurations (\ref{eq:sec3_condition2}) when $\phi =0$, to finite numbers of equations of motion of the supergravities in \cite{Honda:2020sbl, Honda:2021rcd}. Then, it is natural to expect to be able to derive the path-integral of perturbative strings on the string backgrounds by considering fluctuations around (\ref{eq:sec3_condition2}) in string geometry theory.

Because   $\frac{\partial}{\partial \bar{\tau}}$ is a partial derivative in the action, the other coordinates $\bar{\bold{E}}$, $\bm X_{G}^{\mu}$ and $\bm X_{LG}^A$ are fixed when it acts on fields. Thus, 
 $\frac{\partial}{\partial \bar{\tau}}$ does not act on $\bar{\bold{E}}$, $\bm X_{G}^{\mu}$ and $\bm X_{LG}^A$ in the action. This is the same situation as a particle's Lagrangian depending on the time explicitly: the partial derivative with respect to the time act only on the explicit  time dependence on the Lagrangian but does not act on the particle field.
Because the differentials are only with respect to $\bm X_{G}^{\mu}$, $\bm X_{LG}^A$ and $\bar{\tau}$, $\bar{\bold{E}}$ is a constant in the backgrounds (\ref{eq:sec3_condition2}). The dependences of $\bar{\bold{E}}$ on the backgrounds are uniquely determined  by the consistency of the quantum theory of the fluctuations around the backgrounds. Actually, we will find that all the perturbative string amplitudes are derived.  
 The Ricci scalar for this metric is 
\begin{align}
\bar{\bm R} 
&= \bm R - 2\int d\bar\sigma d^2\bar\theta\hat{\bm E}
 \int d\bar\sigma' d\bar\theta'\hat{\bm E}'
 \bm G^{(\mu\bar\sigma\bar\theta)(\mu'\bar\sigma'\bar\theta')}
 (\bm\partial_{(\mu\bar\sigma\bar\theta)}\phi\bm\partial_{(\mu'\bar\sigma'\bar\theta')}\phi
 + \bm\nabla_{(\mu\bar\sigma\bar\theta)}\bm\nabla_{(\mu'\bar\sigma'\bar\theta')}\phi)\nonumber\\
&\quad
- 2\int d\bar\sigma d\bar\theta^{-}\bar e
 \int d\bar\sigma' d^2\bar\theta'\hat{\bm E}'
 \bm G^{(A\bar\sigma\bar\theta^{-})(A'\bar\sigma'\bar\theta'^{-})}
 \bm\partial_{(A\bar\sigma\bar\theta^{-})}\phi\bm\partial_{(A'\bar\sigma'\bar\theta'^{-})}\phi,
\end{align}
where $\bm R$ and $\bm\nabla_{(\mu\bar\sigma\bar\theta)}$ denote the Ricci scalar and the covariant derivative for the metric 
$\bm G_{(\mu\bar\sigma\bar\theta)(\mu'\bar\sigma'\bar\theta')}$, respectively. 
We should note that the term including 
$\Big(\dfrac{\partial}{\partial \bm X_{LG}^A}\Big)^2$ vanishes because the applied functional needs to be proportinal to 
$(\bm X_{LG}^A)^2 = (\bar\theta^{-} \lambda_{G}^A)^2=0$.

Next, we consider fluctuations around these backgrounds.
We only consider fluctuations of metric $\bm h_{IJ}$:
\begin{equation}
\bm{\mathcal G}_{IJ} = \bar{\bm G}_{IJ} + \bm h_{IJ}.
\label{mathcalG}
\end{equation}
The degree of freedom of  perturbative superstrings is identified in \cite{Sato:2017qhj, Sato:2019cno, Sato:2020szq, Sato:2022owj, Sato:2022brv, Nagasaki:2023fnz} with the scalar fluctuation $\mb{\psi}_{dd}$, where 
\begin{equation}
\bm\psi_{IJ}\coloneqq \bm h_{IJ} - \frac12\bar{\bm G}^{I'J'}\bm h_{I'J'} \bar{\bm G}_{IJ}.
\end{equation}
  Thus, we consider only $\mb{\psi}_{dd}$,
namely we set 
\begin{equation}
\bm\psi_{d (\mu\bar\sigma\bar\theta)} = \bm\psi_{(\mu\bar\sigma\bar\theta)(\mu'\bar\sigma'\bar\theta')}=
\bm\psi_{d (A\bar\sigma\bar\theta^{-})}=
\bm\psi_{ (A\bar\sigma\bar\theta^{-}) (A'\bar\sigma'\bar\theta^{'-})}=
\bm\psi_{(\mu\bar\sigma\bar\theta) (A\bar\sigma\bar'\theta^{-})}
= 0.
\end{equation}

In order to obtain a propagator, we add a gauge fixing term corresponding to 
the harmonic gauge to the action \eqref{eq:sec3_adction}  and obtain
\begin{equation}\label{eq:sec3_action_1}
S = \int\mathcal D\tau\mathcal D\bm E\mathcal D\bm X_G \mathcal D\bm X_{LG} \sqrt{-\bm{\mathcal G}}\Big[
 e^{-2\bm\varPhi}\Big(\bm{\mathcal R} + 4\bm\nabla_I\bm\varPhi\bm\nabla^I\bm\varPhi
 - \frac12|\tilde{\bm{\mathcal H}}|^2 - \frac{\alpha'}{4}\mathrm{tr}|\bm{\mathcal F}|^2\Big)
 - \frac12\bm{\mathcal G}^{IJ}(\bm\nabla^{I'}\bm\psi_{I'I})(\bm\nabla^{J'}\bm\psi_{J'J})\Big],
\end{equation}
where we abbreviate the  Faddeev-Popov ghost term because it does not contribute to the ``tree''-level two-point correlation functions of the metrics. 
By considering the action up to the second order in $\bm\psi_{dd}$,
substituting \eqref{mathcalG}, \eqref{eq:sec3_backgrounds_B}, \eqref{eq:sec3_backgrounds_Phi} and \eqref{eq:sec3_backgrounds_C}, 
and taking the limit $D\rightarrow\infty$ after a rather long calculation, 
this becomes 
\begin{align}\label{eq:sec3_action_2}
S &= \int\mathcal D\bar\tau \mathcal D\bm E\mathcal D\bm X_{G}\mathcal D\bm X_{LG}
 \sqrt{-\bm G}e^\phi
 \Big[e^{-2\bm\Phi}\Big(\bm R
 - 2\int d\bar\sigma d\bar\theta\hat{\bm E}
 (\bm\nabla^{(\mu\bar\sigma\bar\theta)}\bm\nabla_{(\mu\bar\sigma\bar\theta)}\phi
 + \bm\partial_{(\mu\bar\sigma\bar\theta)}\phi\bm\partial^{(\mu\bar\sigma\bar\theta)}\phi)\nonumber\\
&\hspace{15ex}  
 - 2\int d\bar\sigma d\bar\theta^{-}\bar e
 (\bm\partial^{(A\bar\sigma\bar\theta^{-})}\bm\partial_{(A\bar\sigma\bar\theta^{-})}\phi
 + \bm\partial_{(A\bar\sigma\bar\theta^{-})}\phi\bm\partial^{(A\bar\sigma\bar\theta^{-})}\phi)\nonumber\\
&\hspace{15ex}  
 + 4\int d\bar\sigma d\bar\theta\hat{\bm E}
 \bm\partial_{(\mu\bar\sigma\bar\theta)}\bm\Phi
 \bm\partial^{(\mu\bar\sigma\bar\theta)}\bm\Phi
 +4\int d\bar\sigma d\bar\theta^{-}\bar e
 \bm\partial_{(A\bar\sigma\bar\theta^{-})}\bm\Phi
 \bm\partial^{(A\bar\sigma\bar\theta^{-})}\bm\Phi
 - \frac12 |\tilde{\bm H}|^2 - \frac{\alpha'}{4}\mathrm{tr}|\bm F|^2\Big)\nonumber\\
&\qquad
 +e^{-2\bm\Phi - 2\phi}\Big(
 \int d\bar\sigma d\bar\theta\hat{\bm E}
 \Big(\bm\nabla^{(\mu\bar\sigma\bar\theta)}\bm\nabla_{(\mu\bar\sigma\bar\theta)}\phi
  + \bm\partial_{(\mu\bar\sigma\bar\theta)}\phi\bm\partial^{(\mu\bar\sigma\bar\theta)}\phi
  \Big)\nonumber\\
&\hspace{15ex}
  + \int d\bar\sigma d\bar\theta^{-}\bar e
 \Big(\bm\partial^{(A\bar\sigma\bar\theta^{-})}\bm\partial_{(A\bar\sigma\bar\theta^{-})}\phi
  + \bm\partial_{(A\bar\sigma\bar\theta^{-})}\phi\bm\partial^{(A\bar\sigma\bar\theta^{-})}\phi
 \Big) 
 \Big)\bm\psi_{dd}\nonumber\\
&\qquad 
 + \frac14e^{-2\bm\Phi - 4\phi}\bm\psi_{dd}
 \Big(\int d\bar\sigma d\bar\theta\hat{\bm E}
 \bm\nabla^{(\mu\bar\sigma\bar\theta)}\bm\nabla_{(\mu\bar\sigma\bar\theta)}
  + \int d\bar\sigma d\bar\theta^{-}\bar e
 \bm\partial^{(A\bar\sigma\bar\theta^{-})}\bm\partial_{(A\bar\sigma\bar\theta^{-})}
 +e^{-2\phi}\partial_d^2
 \Big)
 \bm\psi_{dd}\nonumber\\
&\qquad
 + \frac14 e^{-2\bm\Phi - 4\phi}\Big(-\bm R + \frac12 |\tilde{\bm H}|^2 + \frac{\alpha'}{4}\mathrm{tr}|\bm F|^2\nonumber\\
&\hspace{15ex}
 + \int d\bar\sigma d\bar\theta\hat{\bm E}
 \Big(- \frac12\bm\nabla^{(\mu\bar\sigma\bar\theta)}\bm\nabla_{(\mu\bar\sigma\bar\theta)}\phi
 - \frac{5}2\bm\partial_{(\mu\bar\sigma\bar\theta)}\phi\bm\partial^{(\mu\bar\sigma\bar\theta)}\phi
  - 3\bm\nabla^{(\mu\bar\sigma\bar\theta)}\bm\nabla_{(\mu\bar\sigma\bar\theta)}\bm\Phi\nonumber\\
&\hspace{20ex}
  + 2\bm\partial_{(\mu\bar\sigma\bar\theta)}\bm\Phi
  \bm\partial^{(\mu\bar\sigma\bar\theta)}\bm\Phi
  -10\bm\partial_{(\mu\bar\sigma\bar\theta)}\bm\Phi
 \bm\partial^{(\mu\bar\sigma\bar\theta)}\phi\Big)\nonumber\\
&\hspace{15ex}
 + \int d\bar\sigma d\bar\theta^{-}\bar e
 \Big(-\frac12\bm\partial^{(A\bar\sigma\bar\theta^{-})}\bm\partial_{(A\bar\sigma\bar\theta^{-})}\phi
 - \frac{5}2\bm\partial_{(A\bar\sigma\bar\theta^{-})}\phi\bm\partial^{(A\bar\sigma\bar\theta^{-})}\phi
  - 3\bm\partial^{(A\bar\sigma\bar\theta^{-})}\bm\partial_{(A\bar\sigma\bar\theta^{-})}\bm\Phi\nonumber\\
&\hspace{20ex}
  + 2\bm\partial_{(A\bar\sigma\bar\theta^{-})}\bm\Phi
  \bm\partial^{(A\bar\sigma\bar\theta^{-})}\bm\Phi
  -10\bm\partial_{(A\bar\sigma\bar\theta^{-})}\bm\Phi
 \bm\partial^{(A\bar\sigma\bar\theta^{-})}\phi\Big)\Big)
\bm\psi_{dd}^2\Big].
\end{align}

By normalizing  the kinetic term of $\bm\psi_{dd}$ as 
\begin{equation}\label{eq:sec3_normalize_psi_dd}
\bm\psi_{dd}= 2e^{\bm\varPhi + (3/2)\phi}\bm\psi_{dd}',  
\end{equation}
the action becomes 
\begin{align}\label{eq:sec3_action_3}
S &= \int\mathcal D\bar\tau\mathcal D\bm E\mathcal D\bm X_{G}\mathcal D\bm X_{LG}\sqrt{-\bm G}
\Big[e^{-2\bm\Phi-\phi}
 \Big(\bm R - \frac12|\tilde{\bm H}|^2 - \frac{\alpha'}{4}\mathrm{tr}|\bm F|^2\nonumber\\
&\quad
 - \int d\bar\sigma d\bar\theta\hat{\bm E}
 \big(2\bm\nabla^{(\mu\bar\sigma\bar\theta)}\bm\nabla_{(\mu\bar\sigma\bar\theta)}\phi
  + 2\bm\partial_{(\mu\bar\sigma\bar\theta)}\phi\bm\partial^{(\mu\bar\sigma\bar\theta)}\phi
  - 4\bm\partial_{(\mu\bar\sigma\bar\theta)}\bm\Phi
  \bm\partial^{(\mu\bar\sigma\bar\theta)}\bm\Phi\big)\nonumber\\
&\quad
 - \int d\bar\sigma d\bar\theta^{-}\bar e
 \big(2\bm\partial_{(A\bar\sigma\bar\theta^{-})}\phi\bm\partial^{(A\bar\sigma\bar\theta^{-})}\phi
  - 4\bm\partial_{(A\bar\sigma\bar\theta^{-})}\bm\Phi\bm\partial^{(A\bar\sigma\bar\theta^{-})}\bm\Phi\big)
 \Big)\nonumber\\
&\quad
 +L_1\bm\psi_{dd}' 
 +\int d\bar\sigma d\bar\theta\hat{\bm E}
 \bm\psi_{dd}'\bm\nabla^{(\mu\bar\sigma\bar\theta)}\bm\nabla_{(\mu\bar\sigma\bar\theta)}
 \bm\psi_{dd}'
  + e^{-2\phi} \bm\psi_{dd}' \partial_d^2 \bm\psi_{dd}'
+ L_2\bm\psi_{dd}'^2\Big],
\end{align}
where
\begin{subequations}\label{eq:sec3_def_lagrangian_L1_L2}
\begin{align}
L_1 
&\coloneqq 2e^{-\bm\Phi + \phi/2}
 \Big(\int d\bar\sigma d\bar\theta\hat{\bm E}
 (\bm\nabla^{(\mu\bar\sigma\bar\theta)}\bm\nabla_{(\mu\bar\sigma\bar\theta)}\phi 
  + \bm\partial_{(\mu\bar\sigma\bar\theta)}\phi\bm\partial^{(\mu\bar\sigma\bar\theta)}\phi)
 + \int d\bar\sigma d\bar\theta^{-}\bar e
 \bm\partial_{(A\bar\sigma\bar\theta^{-})}\phi\bm\partial^{(A\bar\sigma\bar\theta^{-})}\phi\Big),\\
L_2 
&\coloneqq -\bm R + \frac12 |\tilde{\bm H}|^2 + \frac{\alpha'}{4}\mathrm{tr}|\bm F|^2
+ \int d\bar\sigma d\bar\theta\hat{\bm E}
 \Big(-\frac12\bm\nabla^{(\mu\bar\sigma\bar\theta)}\bm\nabla_{(\mu\bar\sigma\bar\theta)}\phi
- \frac{1}{4}\bm\partial_{(\mu\bar\sigma\bar\theta)}\phi
  \bm\partial^{(\mu\bar\sigma\bar\theta)}\phi\nonumber\\
&\hspace{5ex}
 - 3\bm\nabla^{(\mu\bar\sigma\bar\theta)}\bm\nabla_{(\mu\bar\sigma\bar\theta)}\bm\Phi
  + 3\bm\partial_{(\mu\bar\sigma\bar\theta)}\bm\Phi\bm\partial^{(\mu\bar\sigma\bar\theta)}\bm\Phi
 -7\bm\partial^{(\mu\bar\sigma\bar\theta)}\bm\Phi
 \bm\partial_{(\mu\bar\sigma\bar\theta)}\phi\Big)\nonumber\\
&\hspace{5ex}
+ \int d\bar\sigma d\bar\theta^{-}\bar e
 \Big(-\frac{1}{4}\bm\partial_{(A\bar\sigma\bar\theta^{-})}\phi
  \bm\partial^{(A\bar\sigma\bar\theta^{-})}\phi
 + 3\bm\partial_{(A\bar\sigma\bar\theta^{-})}\bm\Phi\bm\partial^{(A\bar\sigma\bar\theta^{-})}\bm\Phi
 -7\bm\partial^{(A\bar\sigma\bar\theta^{-})}\bm\Phi
 \bm\partial_{(A\bar\sigma\bar\theta^{-})}\phi\Big). 
\end{align}
\end{subequations}

In order to set a background $\bar{\bm G}_{dd}=e^{2\phi}$, which corresponds to $\bm \psi_{dd}$, on-shell, we shift $\bm\psi_{dd}'$ as 
\begin{equation}
\bm\psi_{dd}'  = \bm\psi_{dd}'' - f, \label{shift}
\end{equation}
by a functional 
\footnote{This field redefinition is local with respect to fields in string geometry theory, because the fields are functionals of the coordinates, $\bm X^\mu_{\hat{D}_{G}} (\bar\sigma, \bar\tau, \bar\theta)$ and $\bm X^A_{LG}(\bar\sigma, \bar\tau, \bar\theta^-)$.
} $f$ of the coordinates $\bm X^\mu_{G}(\bar\sigma, \bar\tau, \bar\theta)$ and $\bm X^A_{LG}(\bar\sigma, \bar\tau, \bar\theta^-)$, so that the first order terms in  $\bm\psi_{dd}''$  vanish. 
This condition is written as 
\begin{equation}\label{eq:sec3_f_diffeq}
\int d\bar\sigma d\bar\theta\hat{\bm E}
 \bm\nabla^{(\mu\bar\sigma\bar\theta)}\bm\nabla_{(\mu\bar\sigma\bar\theta)}f 
 + L_2f = \frac12L_1.
\end{equation}
$f$ exists because this is a second order differential equation for $f$. 
As a result, 
\begin{align}\label{eq:sec3_action_4}
S &= \int\mathcal D\bar\tau\mathcal D\bm E\mathcal D\bm X_{G}\mathcal D\bm X_{LG}
\sqrt{-\bm G}\Big[e^{-2\bm\Phi+\phi}\Big(
 +\bm R - \frac12 |\tilde{\bm H}|^2 - \frac{\alpha'}{4}\mathrm{tr}|\bm F|^2\nonumber\\
&\quad
 - \int d\bar\sigma d\bar\theta\hat{\bm E}
 (-4\bm\partial_{(\mu\bar\sigma\bar\theta)}\bm\Phi
  \bm\partial^{(\mu\bar\sigma\bar\theta)}\bm\Phi
 + 2\bm\nabla^{(\mu\bar\sigma\bar\theta)}\bm\nabla_{(\mu\bar\sigma\bar\theta)}\phi
 + 2\bm\partial_{(\mu\bar\sigma\bar\theta)}\phi\bm\partial^{(\mu\bar\sigma\bar\theta)}\phi)\Big)\nonumber\\
&\quad
 - \int d\bar\sigma d\bar\theta^{-}\bar e
 (-4\bm\partial_{(A\bar\sigma\bar\theta^{-})}\bm\Phi
  \bm\partial^{(A\bar\sigma\bar\theta^{-})}\bm\Phi
 + 2\bm\partial_{(A\bar\sigma\bar\theta^{-})}\phi\bm\partial^{(A\bar\sigma\bar\theta^{-})}\phi)\Big)\nonumber\\
&\quad
 - e^{-\bm\Phi + \phi/2}
 \Big(\int d\bar\sigma d\bar\theta\hat{\bm E}
 (\bm\nabla^{(\mu\bar\sigma\bar\theta)}\bm\nabla_{(\mu\bar\sigma\bar\theta)}\phi
 + \bm\partial_{(\mu\bar\sigma\bar\theta)}\phi\bm\partial^{(\mu\bar\sigma\bar\theta)}\phi)
 + \int d\bar\sigma d\bar\theta^{-}\bar e
 \bm\partial_{(A\bar\sigma\bar\theta^{-})}\phi\bm\partial^{(A\bar\sigma\bar\theta^{-})}\phi\Big)f\nonumber\\
&\quad
 +\bm\psi_{dd}''\int d\bar\sigma d\bar\theta\hat{\bm E}
 \bm\nabla^{(\mu\bar\sigma\bar\theta)}\bm\nabla_{(\mu\bar\sigma\bar\theta)}\bm\psi_{dd}'' 
   + e^{-2\phi} \bm\psi_{dd}'' \partial_d^2 \bm\psi_{dd}''
 + L_2\tilde{\bm\psi}_{dd}''^2\Big]. 
\end{align}

In the following, we consider only states where the dependence of the fluctuation $\bm\psi_{dd}$ on  $\bm X_{G}$  is local with respect to the indices $(\bar{\sigma}, \bar{\theta})$ as
\begin{equation}
\bm\psi_{dd}''[\bar{\bm E}, \bm X_{G}, \bm X_{LG}] 
= \int d\bar\sigma d\bar\theta\,g(\bm X_{G}(\bar\sigma,\bar\theta))[\bar{\bm E}, \bm X_{LG}], 
\end{equation}
where $g(x)$ is an arbitrary local function, 
and obtain a component representation of \eqref{eq:sec3_action_4}. 
Under the super diffeomorphism transformation of  
$\bar\sigma$ and $\bar\theta$, 
$\displaystyle\int d\bar\sigma d\bar\theta\hat{\bm E}$
is invariant, then 
$(1/\hat{\bm E})\delta(\bar\sigma - \bar\sigma')\delta(\bar\theta - \bar\theta')$ is a scalar.
Furthermore, under the super diffeomorphism transformation where $\bar\theta$ is fixed, that is under the diffeomorphism transformation of $\bar\sigma$,
$\displaystyle\int d\bar\sigma\bar e$
is invariant, then 
$(1/\bar e)\delta(\bar\sigma - \bar\sigma')$ is a scalar in this case.
Thus, under the super diffeomorphism transformation 
where $\bar\sigma$ is fixed, 
$(\bar e/\hat{\bm E})\delta(\bar\theta - \bar\theta')$ is a scalar. 
Hence we have 
\begin{align}
\frac{\partial}{\partial X^\mu(\bar\sigma)}
&= \int d\sigma' d\bar{\theta}'\,\hat{\mb{E}}(\bar\sigma', \bar\theta')
\frac{\partial\bm X_{G}^\nu(\bar\sigma', \bar\theta')}{\partial X^\mu(\bar\sigma)}
\frac{\partial}{\partial\bm X_{G}^\nu(\bar\sigma', \bar\theta')} \nonumber \\
&= \int d\sigma' d\bar\theta'\hat{\bm E}(\bar\sigma', \bar\theta')\,
\frac1{\bar e}\delta^\nu_\mu\delta(\bar\sigma - \bar\sigma')
\frac{\partial}{\partial\bm X_{G}^\nu(\bar\sigma', \bar\theta')}\nonumber \\
&= \int d\bar\theta\frac{\hat{\bm E}(\bar\sigma, \bar\theta)}{\bar e(\bar\sigma)}\,
\frac{\pd}{\pd {\bm X}_{G}^{\mu}(\bar{\sigma}, \bar{\theta})}, 
\end{align}
\begin{align}
\frac{\partial}{\partial\psi^\mu(\bar\sigma)}
&= \int d\sigma' d\bar\theta'\hat{\bm E}(\bar\sigma', \bar\theta')
\frac{\partial\bm X_{G}^\nu(\bar\sigma',\bar\theta')} 
{\pd {\psi}^{\mu}(\bar{\sigma})}
\frac{\partial}{\partial\bm X_{G}^\nu(\bar\sigma', \bar\theta')}\nonumber \\
&= \int d\sigma' d\bar\theta'\hat{\bm E}(\bar\sigma', \bar\theta')
\frac1{\bar e}\delta^\nu_\mu\delta(\bar\sigma - \bar\sigma')\bar\theta'
\frac{\partial}{\partial\bm X_{G}^\nu(\bar\sigma', \bar\theta')}\nonumber \\
&= \int d\bar\theta\frac{\hat{\bm E}(\bar\sigma, \bar\theta)}{\bar e(\bar\sigma)}\bar\theta
\frac{\partial}{\partial\bm X_{G}^\mu(\bar\sigma, \bar\theta)}, 
\end{align}

\begin{align}
\frac{\partial}{\partial X^{\mu}(\bar\sigma)}
\frac{\partial}{\partial X^\nu(\bar\sigma)}
&= \int d\bar\theta'\frac{\hat{\mb{E}}(\bar{\sigma}, \bar{\theta}')}
{\bar{e}(\bar{\sigma})}
\int d\bar\theta\,\frac{\hat{\mb{E}}(\bar{\sigma}, \bar{\theta})}
{\bar{e}(\bar{\sigma})}\,
\frac{\partial}{\partial\bm X_{G}^\mu(\bar\sigma, \bar\theta')} 
\frac{\partial}{\partial\bm X_{G}^\nu(\bar\sigma, \bar\theta)}\nonumber \\
&= \int d\bar\theta'\frac{\hat{\bm E}(\bar\sigma, \bar\theta')}{\bar e(\bar\sigma)}
\int d\bar\theta\frac{\hat{\bm E}(\bar\sigma, \bar\theta)}{\bar e(\bar\sigma)}
\frac{\bar e(\bar\sigma)}{\hat{\bm E}(\bar\sigma, \bar\theta)}
\delta(\bar\theta - \bar\theta')
\frac{\partial}{\partial\bm X_{G}^\mu(\bar\sigma, \bar\theta)} 
\frac{\partial}{\partial\bm X_{G}^\nu(\bar\sigma, \bar\theta)}\nonumber \\
&= \int d\bar\theta\frac{\hat{\bm E}(\bar\sigma, \bar\theta)}{\bar e(\bar\sigma)}\,
\frac{\partial}{\partial\bm X_{G}^\mu(\bar\sigma, \bar\theta)}
\frac{\partial}{\partial\bm X_{G}^\nu(\bar\sigma, \bar\theta)},  
\end{align}

\begin{align}
\frac{\partial}{\partial X^\mu(\bar\sigma)}
\frac{\partial}{\partial\psi^\nu(\bar\sigma)}
&= \int d\bar{\theta}'\frac{\hat{\bm E}(\bar\sigma, \bar\theta')}{\bar e(\bar\sigma)}
\int d^2\bar{\theta}\frac{\hat{\bm E}(\bar\sigma, \bar\theta)}{\bar e(\bar\sigma)}
\frac{\partial}{\partial\bm X_{G}^\mu(\bar\sigma,\bar\theta')} 
\bar{\theta}
\frac{\partial}{\partial\bm X_{G}^\nu(\bar\sigma, \bar\theta)}\nonumber\\
&= \int d\bar\theta'\frac{\hat{\mb{E}}(\bar\sigma, \bar\theta')}{\bar e(\bar\sigma)}\,
\int d\bar\theta\; \bar\theta \frac{\hat{\bm E}(\bar\sigma, \bar\theta)}{\bar e(\bar\sigma)}
\frac{\bar e(\bar\sigma)}{\hat{\bm E}(\bar\sigma, \bar\theta)}
\delta(\bar\theta - \bar\theta')
\frac{\partial}{\partial\bm X_{G}^\mu(\bar\sigma, \bar\theta)} 
\frac{\partial}{\partial\bm X_{G}^\nu(\bar\sigma, \bar\theta)}\nonumber\\
&= \int d\bar\theta\; \bar\theta\frac{\hat{\bm E}(\bar\sigma, \bar\theta)}{\bar e(\bar\sigma)}\,
\frac{\partial}{\partial\bm X_{G}^\mu(\bar\sigma, \bar\theta)}
\frac{\partial}{\partial\bm X_{G}^\nu(\bar\sigma, \bar\theta)}, 
\end{align}
and
\begin{align}
G^{(\mu\bar\sigma)(\mu'\bar\sigma')}(\bm X_{G}(\bar\sigma, \bar\theta))
&= \delta^{\bar\sigma\bar\sigma'}
 \Big(G^{\mu\mu'}(X) + \bar\theta\psi^\rho\pd_\rho G^{\mu\mu'}(X) \Big),\\
\Gamma^{(\mu''\bar\sigma'')}_{(\mu\bar\sigma)(\mu'\bar\sigma')}(\bm X_{G}(\bar\sigma, \bar\theta))
&= \delta^{\bar\sigma\bar\sigma'}\delta^{\bar\sigma\bar\sigma''}
 \Big(\Gamma^{\mu''}_{\mu\mu'}(X) + \bar\theta\psi^\rho\pd_\rho\Gamma^{\mu''}_{\mu\mu'}(X) \Big).
\end{align}
Collecting the above results, the action \eqref{eq:sec3_action_4} becomes 
\begin{align}
S &= \int\mathcal D\bar\tau\mathcal D\bm E\mathcal D\bm X_{G}\mathcal D\bm X_{LG}
 \sqrt{-\bm{G}}
 \Big[e^{-2\bm\Phi+\phi}\Big(
 \bm R - \frac12 |\bm H|^2 - \frac{\alpha'}{4}\mathrm{tr}|\bm F|^2\nonumber\\
&\quad
 - \int d\bar\sigma d\bar\theta\hat{\bm E}
  \big(-4\bm\partial_{(\mu\bar\sigma\bar\theta)}\bm\Phi\bm\partial^{(\mu\bar\sigma\bar\theta)}\bm\Phi
  +2\bm\nabla^{(\mu\bar\sigma\bar\theta)}\bm\nabla_{(\mu\bar\sigma\bar\theta)}\phi
  +2\bm\partial_{(\mu\bar\sigma\bar\theta)}\phi\bm\partial^{(\mu\bar\sigma\bar\theta)}\phi\big)\nonumber\\
&\quad
 - \int d\bar\sigma d\bar\theta^{-}\bar e
  \big(-4\bm\partial_{(A\bar\sigma\bar\theta^{-})}\bm\Phi\bm\partial^{(A\bar\sigma\bar\theta^{-})}\bm\Phi
  +2\bm\partial_{(A\bar\sigma\bar\theta^{-})}\phi\bm\partial^{(A\bar\sigma\bar\theta^{-})}\phi\big)
\Big)\nonumber\\
&\qquad
 - e^{-\bm\Phi+\phi/2}\Big(
 \int d\bar\sigma d\bar\theta\hat{\bm E}\big(
 \bm\nabla^{(\mu\bar\sigma\bar\theta)}\bm\nabla_{(\mu\bar\sigma\bar\theta)}\phi
  + \bm\partial_{(\mu\bar\sigma\bar\theta)}\phi\bm\partial^{(\mu\bar\sigma\bar\theta)}\phi\big)
  + \int d\bar\sigma d\bar\theta^{-}\bar e
  \bm\partial_{(A\bar\sigma\bar\theta^{-})}\phi\bm\partial^{(A\bar\sigma\bar\theta^{-})}\phi\Big)f
 \nonumber\\
&\qquad
 + \bm\psi_{dd}''\nabla^2\bm\psi_{dd}''
 + \bm\psi_{dd}''\psi^\rho\partial_\rho G^{\mu\nu}\nabla_\mu\Big(\frac{\partial\bm\psi_{dd}''}{\partial\psi^\nu}\Big)
    + e^{-2\phi} \bm\psi_{dd}'' \partial_d^2 \bm\psi_{dd}''
 + L_2\bm\psi_{dd}''^2\Big].
\label{eq:sec3_action_5}
\end{align}
We further consider only slowly varying $\bm\psi''_{dd}$, namely we make derivative expansions:
\begin{equation}
 \partial_d \bm\psi_{dd}'' \rightarrow \sqrt\epsilon  \partial_d \bm\psi_{dd}'',\quad
\nabla_\mu\bm\psi_{dd}'' \rightarrow \sqrt\epsilon\nabla_\mu\bm\psi_{dd}'',\quad
\frac{\partial\bm\psi_{dd}''}{\partial\psi^\mu} \rightarrow \sqrt\epsilon\frac{\partial\bm\psi_{dd}''}{\partial\psi^\mu},\quad
\bm\psi_{dd}'' \rightarrow \bm\psi_{dd}'',
\label{slowlyvarying}
\end{equation}
where $\epsilon$  is an infinitesimal parameter. This corresponds to a Newtonian limit \cite{Sato:2020szq}.

The second order part of the action can be written as
\begin{equation}
S^{(2)} = -2 \int\mathcal D\bar\tau\mathcal D\bm E\mathcal D\bm X_{G}\mathcal D\bm X_{LG}\bm\psi_{dd}''\sqrt{-G}
 H\Big(-i \frac{\partial}{\partial \bar{\tau}}, -i\frac1{\bar e}\nabla,\frac{\partial}{\partial\psi}, \bm X,\lambda_G,\bar{\bm E}\Big)\bm\psi_{dd}'',
\end{equation}
where
\begin{align}
&H\Big(p_{\bar{\tau}}, p_X,\frac{\partial}{\partial\psi},\bm X,\lambda_G,\bar{\bm E}\Big)\nonumber\\
&= \frac{\epsilon}{2}\int d\bar\sigma\sqrt{\bar h}\Big(
 G^{\mu\nu}{p_X}_\mu{p_X}_\nu
 + i\frac1{\bar e}\psi^\rho\partial_\rho G^{\mu\nu}{p_X}_\mu\Big(\frac{\partial}{\partial\psi^{\nu}}\Big)\Big)
+\frac{\epsilon}{2} e^{-2\phi} p_{\bar{\tau}}^2
 - \frac12L^{(2)}\nonumber\\
&\quad
 + \epsilon\int d\bar\sigma n^{\bar\sigma}\Big(i\bar\partial_{\bar\sigma}X^\mu
  + \frac{\sqrt{\bar h}}{\bar e^2}\partial_{\bar\sigma}X^\nu B_\nu^{\;\;\mu}\Big)
 \bar e{p_X}_\mu
 - \frac{1}2\epsilon\int d\bar\sigma\frac{\bar h}{\bar e^2}\bar E^0_{\bar z}
 \psi^\mu\chi_z\bar e{p_X}_\mu\nonumber\\
&\quad
 -\frac{i}{2} \epsilon\int d\bar\sigma n^{\bar\sigma}
  \frac{\sqrt{\bar h}}{\bar e^2}\partial_{\bar\sigma}X^\nu \nabla_{\mu}B_\nu^{\;\;\mu}
\nonumber\\
&\quad
 + \frac{i}4\epsilon\int d\bar\sigma\frac{\bar h}{\bar e^2}\bar E^0_z\psi^\mu H^{\:\:\nu}_{\mu\:\:\rho}\psi^\rho\bar e{p_X}_\nu
 + \frac18\epsilon\int d\bar\sigma\frac{\bar h}{\bar e^2}\bar E^0_z\psi^\mu
 \nabla_\nu H^{\:\:\nu}_{\mu\:\:\rho}\psi^\rho
 - \frac{i}2\epsilon\int d\bar\sigma\frac{\bar h}{\bar e^2}\bar E^0_z\lambda_G^AA^\mu_{AB}\lambda_G^B\bar e{p_X}_\mu\Big),
\end{align}
and we have added the following total derivative  terms into the action: 
\begin{align}\label{eq:sec3_action_totderterms}
0 &= -2\epsilon\int\mathcal D\bar\tau\mathcal D\bm E\mathcal D\bm X_{G}\mathcal D\bm X_{LG}\sqrt{-G}
\bm\psi_{dd}''\Big[\int d\bar\sigma\bar n^{\bar\sigma}\partial_{\bar\sigma}X^\mu\nabla_\mu
 -i \int d\bar\sigma\frac{\sqrt{\bar h}}{\bar e^2}n^{\bar\sigma}\partial_{\bar\sigma}X^{\rho} B_{\rho}^{\;\;\mu}\nabla_\mu\nonumber\\
&\quad
 + \frac{i}2\int d\bar\sigma\frac{\bar h}{\bar e^2}\bar E^0_{\bar z}\psi^\mu\chi_z\nabla_\mu
  -\frac{i}{2} \int d\bar\sigma n^{\bar\sigma}
  \frac{\sqrt{\bar h}}{\bar e^2}\partial_{\bar\sigma}X^\nu \nabla_{\mu}B_\nu^{\;\;\mu}
 + \frac{1}4\int d\bar\sigma\frac{\bar h}{\bar e^2}\bar E^0_z\psi^\mu H^{\:\:\nu}_{\mu\:\:\rho}\psi^\rho\nabla_\nu\nonumber\\
&\quad
 + \frac{1}8\int d\bar\sigma\frac{\sqrt{\bar h}}{\bar e^2}\bar E^0_z\psi^\mu\nabla_\nu H^{\:\:\nu}_{\mu\:\:\rho}\psi^\rho
 - \frac{1}2\int d\bar\sigma\frac{\bar h}{\bar e^2}\bar E^0_z\lambda_G^AA^\mu_{AB}\lambda_G^B\nabla_\mu\Big]\bm\psi_{dd}''.
\end{align}
The propagator for $\bm\psi_{dd}''$ defined by
\begin{align}
&\Delta_F\big(\bar{\bm E}, \bm X_{G}(\bar\tau),\lambda_{G}(\bar\tau), \bar\tau; 
 \bar{\bm E}', \bm X'_{G}(\bar\tau'), \lambda_{G}(\bar\tau'), \bar\tau' \big)\nonumber\\
&= \big<\bm\psi_{dd}''(\bar{\bm E}, \bm X_{G}(\bar\tau), \lambda_{G}(\bar\tau), \bar\tau),\bm\psi_{dd}''(\bar{\bm E}',\bm X'_{G}(\bar\tau'),\lambda_{G}(\bar\tau'), \bar\tau')\big>,
\end{align}
satisfies
\begin{align}\label{eq:sec3_H_deltaF}
&H\Big(-i \frac{\partial}{\partial \bar{\tau}},-i\frac1{\bar e}\nabla, \frac{\partial}{\pd \psi(\bar{\tau})}, \bm X_{\hat{D}_{G}}(\bar\tau),\lambda_{G}(\bar\tau),\bar{\bm E}\Big)
\Delta_F\big(\bar{\bm E}, \bm X_{G}(\bar\tau),\lambda_{G}(\bar\tau), \bar\tau; 
 \bar{\bm E}', \bm X'_{G}(\bar\tau'), \lambda_{G}(\bar\tau'), \bar\tau \big)\nonumber\\
&= \delta(\bar{\bm E} - \bar{\bm E}')\delta(\bm X_{G}(\bar\tau) - \bm X_{G}'(\bar\tau'))
 \delta(\lambda_{G}(\bar\tau) - \lambda_{G}'(\bar\tau')) \delta(\bar\tau-\bar\tau'). 
\end{align}

In order to obtain a Schwinger representation of the propagator, we use the operator formalism 
$(\hat{\bar{\bm E}}, \hat{\bm X}_{G}(\bar\tau),\hat\lambda_{G}(\bar\tau), \hat{\bar{\tau}})$ of the first quantization. 
The eigenstate for 
$(\hat{\bar{\bm E}},\hat{ X}(\bar\tau), \hat{\bar{\tau}})$ 
is given by $\left|\bar{\bm E}, X(\bar\tau),  \bar{\tau} \right>$.
The conjugate momentum is written as $(\hat{\bm p}_{\bar{\bm E}}, \hat p_X, \hat p_{\bar{\tau}})$. 
The Majorana fermions $\psi^{\mu}$ and $\lambda_{G}^A$ are self-conjugate. Renormalized operators 
$\hat{\tilde{\psi}}^{\mu} \coloneqq \sqrt{\bar{E}_z^0}\,\hat{\psi}^{\mu}$ and 
$\hat{\tilde{\lambda}}_{G}^{A} \coloneqq \sqrt{\bar{E}_{\bar{z}}^0}\,\hat{\lambda}_{G}^{A}$ satisfy $\{ \hat{\tilde{\psi}}^{\mu}(\bar{\sigma}), \hat{\tilde{\psi}}^{\nu}(\bar{\sigma}')\} = \bar{h}^{-1/2}\eta^{\mu \nu} \delta(\bar{\sigma}-\bar{\sigma}')$ and 
$\{ \hat{\tilde\lambda}_{G}^A(\bar\sigma), \hat{\tilde\lambda}_{G}^B(\bar{\sigma}')\} 
 = \bar{h}^{-1/2}\delta^{AB}\delta(\bar\sigma - \bar\sigma')$, respectively.
When we define creation and annihilation operators for 
$\hat{\tilde{\psi}}^\mu$ as $\hat{\hat\psi}^{\hat\mu\dagger} 
 \coloneqq 2^{-1/2}(\hat{\tilde\psi}^{\hat\mu} - i\hat{\tilde\psi}^{\hat\mu + \frac{d}{2}})$ and 
 $\hat{\hat\psi}^{\hat\mu} \coloneqq 2^{-1/2}(\hat{\tilde\psi}^{\hat\mu} + i\hat{\tilde\psi}^{\hat\mu + \frac{d}{2}})$ where $\hat{\mu}=0, \cdots d/2-1$, we have an algebra $\{\hat{\hat{\psi}}^{\hat\mu}(\bar\sigma), \hat{\hat{\psi}}^{\hat\nu\dagger}(\bar\sigma')\} 
 = \bar h^{-1/2}\eta^{\hat\mu\hat\nu} \delta(\bar\sigma - \bar\sigma')$, 
$\{ \hat{\hat{\psi}}^{\hat\mu}(\bar\sigma), \hat{\hat{\psi}}^{\hat\nu}(\bar\sigma')\} = 0$, and 
$\{ \hat{\hat{\psi}}^{\hat\mu\dagger}(\bar\sigma), \hat{\hat{\psi}}^{\hat\nu\dagger}(\bar\sigma')\} = 0$. 
The vacuum $|0\rangle$ for this algebra is defined by $\hat{\hat{\psi}}^{\hat\mu}(\bar\sigma)|0\rangle = 0$. 
The eigen state $|\tilde{\psi}\rangle$, which satisfies 
$\hat{\hat{\psi}}^{\hat\mu}(\bar\sigma)|\tilde\psi\rangle = \tilde\psi^{\hat\mu}(\bar\sigma)|\tilde\psi\rangle$, is given by 
$e^{-\tilde{\psi} \cdot \hat{\hat{\psi}}^{\dagger}} |0\rangle =
e^{- \int d\bar{\sigma} \sqrt{\bar{h}} \tilde{\psi}_{\hat{\mu}}(\bar{\sigma}) \hat{\hat{\psi}}^{\hat{\mu} \dagger}(\bar{\sigma})} |0\rangle$. 
Then, the inner product is given by $\langle\tilde{\psi} | \tilde{\psi}'\rangle = e^{\tilde{\psi}^\dagger \cdot \tilde{\psi}'}$, while the completeness relation is
$\displaystyle\int\mathcal D\tilde\psi^\dagger\mathcal{D}\tilde{\psi} |\tilde\psi\rangle e^{-\tilde{\psi}^\dagger \cdot \tilde{\psi}}\langle\tilde{\psi}| = 1$. 
The same is applied to $\hat{\tilde{\lambda}}_{G}^A(\bar\tau)$. 

Since \eqref{eq:sec3_H_deltaF} means that $\Delta_F$ is an inverse of $H$, $\Delta_F$ can be expressed by a matrix element of the operator $\hat{H}^{-1}$ as
\begin{align}\label{eq:sec3_inverseH}
&\quad
\Delta_F(\bar{\bm E}, \bm X_{G}(\bar\tau), \lambda_{G}(\bar\tau), \bar{\tau}; 
 \bar{\bm E}', \bm X_{G}'(\bar\tau'), \lambda_{G}'(\bar\tau'), \bar{\tau}') = \nonumber\\
&\quad
\langle\bar{\bm E}, \bm X_{G}(\bar\tau), \lambda_{G}(\bar\tau), \bar{\tau}| 
 H^{-1}(\hat p_{\bar{\tau}}, \hat p_X(\bar\tau), \sqrt{\hat{\bar h}}\hat G_{\mu\nu}\hat\psi^\nu(\bar\tau),\hat{\bm X}_{G}(\bar\tau),\hat\lambda_{G}(\bar\tau), \hat{\bar{\bm E}}) |\bar{\bm E}', \bm X_{G}'(\bar\tau'), \lambda_{G}'(\bar\tau'), \bar{\tau}' \rangle.
\end{align}
On the other hand,
\begin{equation}\label{eq:sec3_IntegralFormula}
\hat H^{-1}=  \int _0^\infty dT e^{-T\hat H}, 
\end{equation}
because
\begin{equation}
\lim_{\epsilon\to0+}\int _0^\infty dT e^{-T(\hat H + \epsilon)}
= \lim_{\epsilon\to0+}\bigg[\frac{1}{-(\hat H + \epsilon)} e^{-T(\hat H + \epsilon)}\bigg]_0^\infty
= \hat H^{-1}.
\end{equation}
This fact and \eqref{eq:sec3_inverseH} imply
\begin{align}
&\Delta_F(\bar{\bm E}, \bm X_{G}(\bar\tau), \lambda_{G}(\bar\tau), \bar{\tau}; \bar{\bm E}', \bm X_{G}'(\bar\tau'), \lambda_{G}'(\bar\tau'), \bar{\tau}')\nonumber\\
&\quad
= i\int _0^\infty dT\langle\bar{\bm E}, \bm X_{G}(\bar{\tau}), \lambda_{G}(\bar\tau), \bar{\tau}|  e^{-T\hat H} |\bar{\bm E}', \bm X_{G}'(\bar{\tau}'),  \lambda_{G}'(\bar{\tau}'), \bar{\tau}'\rangle.
\end{align}

In order to define two-point correlation functions that are invariant under the general coordinate transformations in the string geometry, we define in and out states as
\begin{align}
\big\Vert\bm X_{G, i},\lambda_{G,i} \,|\,\bm E_f; \bm E_i\big>_{\rm in} 
 &\coloneqq \int_{\bm E_i}^{\bm E_f} \mathcal D \bm E'
  \big|\bar{\bm E}',\bm X_{G,i},\lambda_{G,i}, \bar\tau' = -\infty \big> \nonumber\\
\big<\bm X_{G, f},\lambda_{G,f} \,|\,\bm E_f ; \bm E_i\big\Vert_{\rm out}
 &\coloneqq \int_{\bm E_i}^{\bm E_f} \mathcal D \bm E
 \big<\bar{\bm E}, \bm X_{G,f},\lambda_{G,f}, \bar\tau = \infty \big|,
\end{align}
where $\bm X_{G,i}\coloneqq \bm X_{G}(\bar\tau' = -\infty)$, 
$\bm X_{G,f}\coloneqq \bm X_{G}(\bar\tau = \infty)$, and $\bm E_i$ and $\bm E_f$ represent the vielbeins of the super cylinders at $\bar\tau = \pm\infty$, respectively. 
$\displaystyle\int$ in $\displaystyle\int\mathcal D\bm E$ includes 
$\sum_{\text{compact}\atop\text{topologies}}$, where
$\mathcal D\bm E$ is the invariant measure
of the super vielbein $\bm E$ on the two-dimensional super Riemannian manifolds $\bm\Sigma$.
$\bm E$ and $\bar{\bm E}$ are related to each others by the super diffeomorphism and the super Weyl transformations. 
When we insert asymptotic states, we integrate out $\bm X_{G,f}$, $\bm X_{G,i}$, $\lambda_{G,f}$, $\lambda_{G,i}$, $\bm E_f$ and $\bm E_i$ in the two-point correlation function for these states;
\begin{align}\label{eq:sec3_def_DeltaF}
&\Delta_F(\bm X_{G,f}; \bm X_{G,i},\lambda_{G,f};\lambda_{G,i}|\bm E_f; \bm E_i)\nonumber\\
&\coloneqq \int_0^\infty dT\big\langle\bm X_{G,f},\lambda_{G,f}|\bm E_f ; \bm E_i\big\Vert_{\rm out}  
 e^{-T\hat H}\big\Vert\bm X_{G,i}\lambda_{G,i}\,|\,\bm E_f ; \bm E_i\big\rangle_{\rm in}.
\end{align}
$\Delta_F(\bm X_{G,f}; \bm X_{G,i},\lambda_{G,f};\lambda_{G,i}|\bm E_f ; \bm E_i)  $ can be written in a path integral representation because it is a time evolution of the states as in \eqref{eq:sec3_def_DeltaF},
\begin{align}\label{eq:sec3_DeltaF_1}
&\Delta_F(\bm X_{G,f}; \bm X_{G,i},\lambda_{G,f};\lambda_{G,i}|\bm E_f ; \bm E_i) \nonumber\\
&= \int_{\bm E_i \bm X_{G,i},\lambda_{G,i}}^{\bm E_f, \bm X_{G,f},\lambda_{G,f}} 
 \mathcal D\bm E\mathcal D\bm X_{G}(\bar\tau)\mathcal D\lambda_{G}(\bar\tau)
  \mathcal{D}\bar{\tau}  \int \mathcal DT\int\mathcal Dp_T\mathcal Dp_X(\bar\tau)\mathcal{D}p_{\bar{\tau}}\nonumber\\
&\quad 
\exp\bigg[-\int_{-\infty}^\infty dt\Big(
 -ip_T(t) \frac{d}{dt}T(t) 
 -i p_{\bar{\tau}}(t)\frac{d}{dt}\bar{\tau}(t)
 - i\int d\bar\sigma\bar e p_{X\mu}(\bar\tau(t), t) \frac{d}{dt} X^{\mu}(\bar\tau(t), t)\nonumber\\
&\qquad 
 + \int d\bar\sigma\frac{i}{2}\sqrt{\bar h}\Big(G_{\mu\nu}(X(\bar\tau(t), t))\,\bar\psi^\mu(\bar\tau(t), t)
\bar E^0_z\frac{d}{dt}\psi^\nu(\bar\tau(t), t)
 + \lambda^A_{G}(\bar\tau(t),t)\bar E^0_z\frac{d}{dt}\lambda_{G,A}(\bar\tau(t),t)\Big)
\nonumber \\
&\qquad 
 + T(t) H\big(p_{\bar{\tau}}(t), p_X(\bar\tau(t), t), \sqrt{\bar h}G_{\mu\nu}(X(\bar\tau(t), t))\,\psi^\nu(\bar\tau(t), t), \bm X_{G}(\bar\tau(t), t), \bar{\bm E}\big)\Big)\bigg].
\end{align}
A derivation in detail is shown in Appendix A.

We also make derivative expansion on the first-quantized fields,
\begin{equation}\label{eq:sec3_L_1}
\frac{d\bar\tau}{dt} \;\rightarrow\; \epsilon\frac{d\bar\tau}{dt},\quad
\frac{dT}{dt} \;\rightarrow\; \epsilon\frac{dT}{dt},\quad
\frac{dX^\mu}{dt} \;\rightarrow\; \epsilon\frac{dX^\mu}{dt},\quad
\frac{d\psi^\mu}{dt} \;\rightarrow\; \epsilon\frac{d\psi^\mu}{dt},\quad
\frac{d\lambda_G^A}{dt} \;\rightarrow\; \epsilon\frac{d\lambda_G^A}{dt},
\end{equation}
so as to be consistent with \eqref{slowlyvarying}, where $\psi_{dd}$ is slowly varying. 
By integrating out $p_{X}(\bar{\tau}(t), t)$ and $p_{\bar{\tau}}(t)$, we move from the canonical formalism to the Lagrange formalism.
Because the exponent of (\ref{eq:sec3_DeltaF_1}) is at most the second order in $p_{X}(\bar{\tau}(t), t)$ and $p_{\bar{\tau}}(t)$, integrating out $p_{X}(\bar{\tau}(t), t)$ and $p_{\bar{\tau}}(t)$ is equivalent to substituting into  (\ref{eq:sec3_DeltaF_1}), the solutions $p_{X}(\bar{\tau}(t), t)$ and $p_{\bar{\tau}}(t)$ of  
\begin{eqnarray}
i \frac{dX^\mu}{dt} - T\Big( \bar e^2G^{\mu\nu}{p_X}_\nu
 + i \bar n^{\bar\sigma}\partial_{\bar\sigma}X^\mu 
  +\frac{\sqrt{\bar h}}{\bar e^2}G^{\mu\nu} Y_\nu\Big)&= &0, \nonumber \\
i \frac{d\bar\tau}{dt} - Te^{-2\phi}p_{\bar\tau} &=& 0,
\end{eqnarray}
where 
\begin{equation}\label{eq:sec3_def_Y}
Y_\mu(\bar\sigma)
= \bar n^{\bar\sigma}\partial_{\bar\sigma}X^\nu B_{\nu\mu} - \frac12\sqrt{\bar h}\bar E^0_z
 \big(\psi^\nu\tilde\Gamma_{\nu\mu\rho}\psi^\rho - i\psi_\mu\chi_z
  + i\lambda_G^AA_{\mu,AB}\lambda_G^B\big),
\end{equation}
which are obtained by differentiating  the exponent of   (\ref{eq:sec3_DeltaF_1})
with respect to $p_{X}(\bar{\tau}(t), t)$ and $ p_{\bar\tau}(t)$, respectively. The solutions are given by  
\begin{align*}
{p_X}_\mu 
&=i \frac{\bar e}{\sqrt{\bar h}}G_{\mu\nu}
 \Big(\frac1T\frac{dX^\nu}{dt} - \bar n^{\bar\sigma}\partial_{\bar\sigma}X^\nu\Big)
 - \frac{1}{\bar e}Y_\mu, \\
 p_{\bar\tau} &= i \frac{e^{2\phi}}{T}\frac{d\bar\tau}{dt}. 
\end{align*}
By substituting this and using the ADM decomposition of the two-dimensional metric,
\begin{equation}\label{eq:def_ADHM_matrix}
\bar h_{mn} 
= \begin{pmatrix}
\bar n^2 + \bar n_{\bar\sigma}\bar n^{\bar\sigma}& \bar n_{\bar\sigma}\\ 
\bar n_{\bar\sigma}& \bar e^2\end{pmatrix},\quad
\bar h^{mn} 
= \begin{pmatrix}
\displaystyle\frac1{\bar n^2}& \displaystyle -\frac{\bar n^{\bar\sigma}}{\bar n^2}\\ 
\displaystyle -\frac{\bar n^{\bar\sigma}}{\bar n^2}& \displaystyle\frac1{\bar e^2} 
 + \Big(\frac{\bar n^{\bar\sigma}}{\bar n}\Big)^2\end{pmatrix},\quad
\bar h = (\bar n\bar e)^2,
\end{equation}
we obtain
\begin{equation}\label{eq:sec3_DeltaF_action}
\Delta_F(\bm X_{G,f}; \bm X_{G,i},\lambda_{G,f};\lambda_{G,i}|\bm E_f ; \bm E_i) 
= \int_{\bm E_i,\bm X_{G,i},\lambda_{G,i}}^{\bm E_f,\bm X_{G,f},\lambda_{G,f}}
 \mathcal DT\mathcal D\bm E\mathcal D\bm X
   \mathcal{D}\bar{\tau}
   \mathcal D\lambda_G(\bar\tau)
 \mathcal Dp_T
 \exp\Big(-\int_{-\infty}^\infty L(t)dt\Big), 
\end{equation}
where 
\begin{align}\label{eq:sec3_Lt}
L(t) &= -i p_T\frac{dT}{dt} 
+ \frac{e^{2\phi}}{2}\frac1{T}\Big(\frac{d\bar\tau}{dt}\Big)^2
 + \frac\epsilon2\int d\bar\sigma\sqrt{\bar h}G_{\mu\nu}
 \Big(\frac1T\bar h^{00}\partial_t X^\mu\partial_t X^\nu
 + 2\bar h^{01}\partial_t X^\mu\partial_{\bar\sigma}X^\nu 
 + T\bar h^{11}\partial_{\bar\sigma}X^\mu\partial_{\bar\sigma}X^\nu\Big)\nonumber\\
&\quad
 + i\epsilon\int d\bar\sigma B_{\mu\nu}\partial_t X^\mu\partial_{\bar\sigma}X^\nu
 + \epsilon\frac{i}{2}\int d\bar\sigma\sqrt{\bar h}G_{\mu\nu}\psi^\mu
 (\bar E^0_z\partial_t + T\bar E^1_z\partial_{\bar\sigma})\psi^\nu
\nonumber\\
&\quad
 + \epsilon\frac{i}{2}\int d\bar\sigma\sqrt{\bar h}
 (\bar E^0_z\partial_t + T\bar E^1_z\partial_{\bar\sigma})X^\mu\psi^\nu\Big(\Gamma_{\nu\mu\rho} - \frac12H_{\nu\mu\rho}\Big)\psi^\rho
 + \frac{1}{2}\epsilon\int d\bar\sigma\sqrt{\bar h}
 (\bar E^0_z\partial_t + T\bar E^1_z\partial_{\bar\sigma})X^\mu\psi_\mu\chi_z\nonumber\\
&\quad
 + \epsilon\frac{i}{2}\int d\bar\sigma\sqrt{\bar h}\lambda_G^A
 (\bar E^0_z\partial_t + T\bar E^1_z\partial_{\bar\sigma})\lambda_{G A}
 - \frac12\epsilon\int d\bar\sigma\sqrt{\bar h}
 (\bar E^0_z\partial_t + T\bar E^1_z\partial_{\bar\sigma})X^\mu\lambda_G^AA_{\mu,AB}\lambda_G^B\nonumber\\
&\quad 
 - \frac{i}{4}\epsilon T\int d\bar\sigma\sqrt{\bar h}F_{\mu\nu,AB}\psi^\mu\psi^\nu\lambda_G^A\lambda_G^B
 +\epsilon \int d\bar\sigma\sqrt{\bar h}\frac{\gamma}{2\pi}TR_{\bar h}.
\end{align}
Here we have fixed a background $\phi$ that satisfies
\begin{align}\label{eq:sec3_phi_equation}
&L_2=L_{\bf GBA}, 
\end{align}
where
\begin{align}\label{eq:sec3_phi_equationWith}
L_{\bf GBA} &\coloneqq   -\epsilon\int d\bar\sigma\sqrt{\bar h}\Big(
 \frac{1}{\bar e^2}G_{\mu\nu}(\partial_{\bar\sigma}X^\mu\partial_{\bar\sigma}X^\nu + Y^\mu Y^\nu)
 -\frac{1}{4\bar e^2}\bar E^0_z\psi^\mu\nabla^\nu H_{\mu\nu\rho}\psi^\rho\nonumber\\
&\qquad
 + (\bar E^0_z\bar n^{\bar\sigma}\partial_{\bar\sigma}X^\mu + \bar E^1_zX'^\mu)
 (i\psi^\nu\tilde\Gamma_{\nu\mu\rho}\psi^\rho + \psi_\mu\chi_z - \lambda_G^AA_{\mu,AB}\lambda_G^B)\nonumber\\
&\qquad
 - \frac{i}{2}F_{\mu\nu,AB}\psi^\mu\psi^\nu\lambda_G^A\lambda_G^B
 + i  \frac{\sqrt{\bar h}}{\bar e^2} n^{\bar\sigma}\partial_{\bar\sigma}X^\nu \nabla_{\mu}B_\nu^{\;\;\mu}
 + \frac{\gamma}{\pi}R_{\bar h}\Big),
\end{align}
where $\gamma$ is an arbitrary constant.
This condition has a consistent $\epsilon$ expansion because in $L_2$,  $\epsilon$ expansion of the backgrounds around the flat background starts at the first order.
This condition is satisfied because it is a second order differential equation for $\phi$ at each order in the $\epsilon$ expansion. 
In this way, $\phi$ can generate all the terms without $\bar{\tau}$ derivatives in the string action as in (\ref{eq:sec3_phi_equation}) with (\ref{eq:sec3_phi_equationWith}), but cannot do those with $\bar{\tau}$ derivatives,   which need  to be derived non-trivially, because the coordinates  $ X^{\mu}(\bar{\tau})$ in string geometry theory are defined on the $\bar{\tau}$ constant lines.

We should note that the time derivative in \eqref{eq:sec3_Lt} is in terms of $t$, not $\bar{\tau}$ at this moment. In Appendix A, we show that $t$ can be fixed to $\bar{\tau}$ by using a reparametrization of $t$ that parametrizes a trajectory. The result is 
\begin{equation}\label{eq:sec3_DeltaF_result}
\Delta_F(\bm X_{G,f}; \bm X_{G,i},\lambda_{G,f};\lambda_{G,i}|\bm E_f ; \bm E_i) 
= Z\int_{\bm E_i,\bm X_{G,i},\lambda_{G,i}}^{\bm E_f,\bm X_{G,f},\lambda_{G,f}}
\mathcal D\bm E\mathcal D\bm X\mathcal D\lambda_G(\bar\tau) e^{-\gamma\chi}e^{-S_{ s}}, 
\end{equation}
where 
\begin{align}\label{eq:sec3_action_result}
S_{\rm s} &= \frac{1}{2\pi \alpha'}\int d\sigma\sqrt{h(\sigma,\tau)}\Big(
 \big(h^{mn}(\sigma,\tau)G_{\mu\nu}(X(\sigma,\tau)) 
  + i\varepsilon^{mn}(\sigma,\tau)B_{\mu\nu}(X(\sigma,\tau))\big)\partial_m X^\mu\partial_n X^\nu\nonumber\\
&\quad
 + iE^a_zG_{\mu\nu}(X(\sigma,\tau))\psi^\mu(\sigma,\tau)\partial_a\psi^\nu(\sigma,\tau)\nonumber\\
&\quad
 + iE^a_z\partial_aX^\mu(\sigma,\tau)\psi^\nu(\sigma,\tau)
 \Big(\Gamma_{\nu\mu\rho} - \frac12H_{\nu\mu\rho}(X(\sigma,\tau))\Big)\psi^\rho(\sigma,\tau)
 + E^a_z\partial_a X^\mu(\sigma,\tau)\psi_\mu(\sigma,\tau)\chi_z(\sigma,\tau)\nonumber\\
&\quad
 + iE^a_z\lambda_G^A\big(\partial_a\lambda_{G A}(\sigma,\tau) 
  + \partial_aX^\nu(\sigma,\tau) A_{\nu,AB}\lambda_G^B(\sigma,\tau)\big)
 - \frac{i}{2}F_{\mu\nu,AB}\psi^\mu(\sigma,\tau)\psi^\nu(\sigma,\tau)
  \lambda_G^A(\sigma,\tau)\lambda_G^B(\sigma,\tau)\Big).
\end{align}
These are the path-integrals of heterotic  perturbative superstrings on arbitrary backgrounds that possess the supermoduli in the SO(32) and $E_8\times E_8$ heterotic superstring theory for $G=$ SO(32) and $E_8\times E_8$, respectively \cite{Brooks:1986uh, Polchinski:1998rr}. 
Therefore, the backgrounds \eqref{eq:sec3_condition2} represents perturbative vacua in heterotic superstring theory.
A consistency of the fluctuation of string geometry, which is the super Weyl invariance  in perturbative superstring theories, implies that the superstring backgrounds are solutions to the equations of motion of the low-energy effective action, that is the heterotic supergravity.  

\section{The potential for heterotic superstring backgrounds}

In this section, we will obtain a potential for heterotic string backgrounds by substituting the heterotic string vacua identified in the previous section into the ``classical'' potential in string geometry theory. One can compare the energies of semi-stable vacua by using this  potential because string geometry theory possesses all order information of  string coupling even in the ``classical'' level of string geometry theory as one can see in the previous sections, and the non-renormalization theorem in string geometry theory states that there is no ``loop'' correction \cite{Sato:2025wfc}.

In the previous section, we derived the path-integrals of heterotic perturbative strings on string backgrounds from the fluctuations around the heterotic string vacua, which include the general heterotic string backgrounds $G_{\mu\nu}(x)$, $B_{\mu\nu}(x)$, $\Phi(x)$, and $A_{\mu}(x)$.  Under the normalization (\ref{eq:sec3_normalize_psi_dd})  and the shift (\ref{shift}) of the fluctuation in this derivation,  the background (\ref{eq:sec3_backgrounds}) and (\ref{eq:sec3_condition2}) becomes 
\begin{equation}
\overline{\overline{G}}_{IJ}=\overline{G}_{IJ}+
2e^{\Phi+\frac{3}{2} \phi}f(- \delta_{I, d}\delta_{J, d}
+\frac{1}{2}e^{-2\phi}\overline{G}_{IJ}),
\end{equation}
which is the explicit form of the string background configurations. By substituting these heterotic string vacua into the ``classical'' potential of string geometry theory, we obtain a potential restricted to the heterotic string background. We call it a potential for heterotic  string backgrounds   $G_{\mu\nu}(x)$, $B_{\mu\nu}(x)$, $\Phi(x)$, and $A_{\mu}(x)$ because it is independent of the string geometry time $\bar{\tau}$ and also the time in the four dimensions when we  impose Poincar$\acute{\mbox{e}}$ invariance in the four dimensions in order to consider the Standard Model of particle physics and its corrections.  
This potential $V_{\rm string}$ is also given by $V_{\rm string}= -S^{(0)}$, where
\begin{align}\label{eq:sec3_S0}
S^{(0)}
&= \int\mathcal D\bar\tau\mathcal D\bm E\mathcal D\bm X_G \mathcal D\bm X_{LG}
 \sqrt{-\bm G}\Big[e^{-2\bm\Phi+\phi}\Big(\bm R 
  - \frac12 |\tilde{\bm H}|^2 - \frac{\alpha'}{4}\mathrm{tr}|\bm F|^2\nonumber\\
&\qquad\qquad
  +4\bm\partial_I\bm\Phi\bm\partial^I\bm\Phi
  - 2\bm\nabla^2\phi
  - 2\bm\partial_I\phi\bm\partial^I\phi\Big)
 - e^{-\bm\Phi+\phi/2}\big(\bm\nabla^2\phi + \bm\partial_I\phi\bm\partial^I\phi\big)f\Big],
\end{align}
which is obtained if the fluctuations are turned off in   (\ref{eq:sec3_action_5}), 
 because $S^{(0)}$ is independent of the string geometry time $\bar{\tau}$. 

The heterotic string backgrounds in the potential must satisfy the equations of motion of the low-energy effective action in heterotic perturbative string theory as stated in the last paragraph of the previous section.  Thus, heterotic perturbative vacua are local minima of the  potential imposed these equations of motion as constraints by the method of Lagrange multiplier. If we fix these local minima and consider fluctuations around them, we can obtain heterotic perturbative strings as in the last section.  
 Furthermore, a non-perturbative correction in string coupling with the order $e^{-1/g_s^2}$ is given by a transition amplitude representing a tunneling process between the semi-stable vacua  in the ``classical'' potential by an ``instanton'' in the theory \cite{Sato:2025wfc}.
From this effect, a generic initial state will reach the minimum of the potential. 
Therefore, the authors in \cite{Nagasaki:2023fnz} conjecture that the ``classical'' potential  restricted to the perturbative vacua in the whole sector of  string geometry theory, called the potential for string backgrounds, represent the string theory landscape and the minimum of the potentials gives the true vacuum in string theory. Especially, $G_{ij}$, which represents the six-dimensional internal space in string theory, will be determined. 

In \eqref{eq:sec3_S0}, 
$\phi$ and $f$ are solutions to \eqref{eq:sec3_phi_equation} and \eqref{eq:sec3_f_diffeq}, respectively. 
Then, by imposing the conditions  \eqref{eq:sec3_phi_equation} and \eqref{eq:sec3_f_diffeq} to \eqref{eq:sec3_S0} by the method of Lagrange multipliers,  we obtain an exact potential,
\begin{align}
V_{\rm string}
=& \int\mathcal D\bar\tau\mathcal D\bm E\mathcal D\bm X_G \mathcal D\bm X_{LG}
 \sqrt{-\bm G} \nonumber \\
 & \Bigg[e^{-2\bm\Phi+\phi}\Big(-\bm R 
  + \frac12 |\tilde{\bm H}|^2 + \frac{\alpha'}{4}\mathrm{tr}|\bm F|^2
  - 4(\bm\partial \bm\Phi)^2 \Big)\nonumber\\
& +(2e^{-2\bm\Phi+\phi}+ f e^{-\bm\Phi+\phi/2})\big(\bm\nabla^2\phi + (\bm\partial \phi)^2\big) \nonumber\\
&+P(L_2-L_{\bf GBA}) \nonumber \\
& +Q \Big( \int d\bar\sigma d\bar\theta\hat{\bm E}
 \bm\nabla^{(\mu\bar\sigma\bar\theta)}\bm\nabla_{(\mu\bar\sigma\bar\theta)}f 
 + L_2f - \frac12L_1 \Big)
 \Bigg],
\end{align}

Here, we take a particle limit, 
$\bold{X}_G^\mu(\bar{\sigma}, \bar{\theta}, \bar{\tau}) \to x^\mu$ on $V_{\rm string}$, which has stringy effects. 
In this limit, string coordinates $X$ reduce to the ten-dimensional coordinates where
\begin{equation}
 \int\mathcal D\bar\tau\mathcal D\bm E\mathcal D\bm X_G \mathcal D\bm X_{LG}
 \sqrt{-\bm G}
\:\to\: \frac{1}{2\kappa_{10}^2}\int d^{10}x \sqrt{-G(x)}. \label{ParticleLimit}
\end{equation}
Thus, $V_\text{string}$ reduces to 
\begin{align}
V_{\rm particle}
=&\frac1{2 \kappa_{10}^2}\int d^{10}x
\sqrt{-G} \nonumber \\
 & \Bigg[e^{-2\Phi+\phi}\Big(-R 
  + \frac12 |\tilde{H}|^2 + \frac{\alpha'}{4} \mathrm{tr}|F|^2 
  - 4(\partial \Phi)^2 \Big)\nonumber\\
&
 + (2e^{-2\Phi+\phi}+f e^{-\Phi+\frac12\phi})\big(\nabla^2\phi + (\partial \phi)^2 \big) \nonumber\\
&+P \Big(\nabla^2\phi
+\frac{1}{2}
  (\partial \phi)^2
+ 14
 \partial^{\mu} \Phi\partial_{\mu}\phi
 + 6  \nabla^2\Phi
 - 6 (\partial \Phi)^2 
  +2R -  |\tilde{H}|^2 -\frac{\alpha'}{2} \mathrm{tr}|F|^2 \Big) \nonumber \\
& +Q \Big(\nabla^2 f -e^{-\Phi+\frac12\phi}
  \Big(\nabla^2 \phi 
   +(\partial \phi)^2\Big)\Big)
 \Bigg]. 
\end{align}

On the other hand, one can obtain an explicit potential without new variables by solving \eqref{eq:sec3_f_diffeq} and \eqref{eq:sec3_phi_equation} and substituting the solutions into \eqref{eq:sec3_S0}. 
In order to search for the minimum of the potential,  one needs to solve \eqref{eq:sec3_f_diffeq} and \eqref{eq:sec3_phi_equation} completely and obtain  exact or series solutions, because the potential is defined globally.  A  potential for searching for the minimum is obtained in \cite{Takeuchi} by  limiting the region to a specific class of string backgrounds, solving \eqref{eq:sec3_f_diffeq} and \eqref{eq:sec3_phi_equation}, and obtaining series solutions.

In order to apply the potential to string phenomenology, we restrict the region of the string backgrounds in the following. 
First, by moving from the string frame to the Einstein frame, where the transformation is given by $G_{\mu\nu} \to e^{\Phi/2}G_{\mu\nu}$,  we obtain 
\begin{align}
V_\text{Einstein}=
&\frac1{2 \kappa_{10}^2}\int d^{10}x
\sqrt{-G} \nonumber \\
 &\Big[e^\phi
 \Big(-R + \frac12e^{-\Phi}|\tilde H|^2 + \frac{\alpha'}{4}e^{-\frac12\Phi}\mathrm{tr}|F|^2
 - \frac12\nabla^2\Phi + \frac12(\partial\Phi)^2\Big)
 \nonumber \\
 &\ 
 + (2e^\phi + fe^{\Phi + \frac12\phi})(\nabla^2\phi + (\partial\phi)^2 + 2\partial_\mu\Phi\partial^\mu\phi)\nonumber\\
 &+P\Big(\nabla^2\phi
+\frac{1}{2}
  (\partial \phi)^2
+ 16
 \partial^{\mu} \Phi\partial_{\mu}\phi
 + 7  \nabla^2\Phi
 -3 (\partial \Phi)^2 
  +2R - e^{-\Phi} |\tilde{H}|^2 -\frac{\alpha'}{2}  e^{-\frac{1}{2}\Phi}\mathrm{tr}|F|^2 \Big) \nonumber \\
& +Q \Big(\nabla^2 f+2\partial_{\mu}f\partial^{\mu}\Phi -e^{-\Phi+\frac12\phi}
  \Big(\nabla^2 \phi 
   +\partial_{\mu}\phi\partial^{\mu}\phi +2\partial_{\mu}\phi\partial^{\mu}\Phi \Big)\Big)
 \Bigg].
\end{align}
Next, by restricting the region to the warped compactification, where 
the metric is given by 
\begin{equation}\label{eq:sec4_metric}
ds^2 = e^{2\rho(y)}\eta_{pq}dx^pdx^q + e^{-2\rho(y)}g_{mn}(y)dy^mdy^n,
\end{equation}
where $p, q= 0, \cdots, 3$ and $m,n= 4, \cdots, 9$,
and the other non-zero backgrounds are given by  $B_{mn}(y)$,  $\Phi(y)$, $A_m(y)$, and $\rho(y)$, we obtain
\begin{align}
V_\text{warp} 
&= \int d^6y\sqrt{g}\Big[
 e^\phi\Big(-R + \frac12e^{-\Phi+4\rho}|\tilde H|^2 +\frac{\alpha'}{4} e^{-\frac12\Phi+2\rho}\mathrm{tr}|F|^2
 - 2\nabla^2\rho + 8(\partial \rho)^2 - \frac12\nabla^2\Phi + \frac12(\partial\Phi)^2\Big) \nonumber \\
& \quad + (2e^\phi + fe^{\Phi+\frac12\phi})(\nabla^2\phi + (\partial\phi)^2 + 2\partial^m\Phi\partial_m\phi)\Big] \nonumber \\
& \quad 
+P\Big(\nabla^2\phi + \frac{1}{2}(\partial\phi)^2
 +16 \partial^m\Phi\partial_m\phi
 + 7\nabla^2\Phi -3(\partial\Phi)^2 \nonumber \\
& \qquad \quad +2R -e^{-\Phi+4\rho}|\tilde H|^2-\frac{\alpha'}{2} e^{-\frac12\Phi+2\rho}\mathrm{tr}|F|^2  +4\nabla^2\rho - 16(\partial \rho)^2  \Big) \nonumber\\
& \quad 
+ Q \Big(\nabla^2 f + 2\partial_m\Phi\partial^m f 
 - e^{-\Phi+\frac12\phi}(\nabla^2\phi + (\partial\phi)^2 + 2\partial^m\Phi\partial_m\phi) \Big).
\label{eq:sec5_diffeq_wcomp_f_1}
\end{align}
The true vacuum in the heterotic sector will be given by a string background that minimize this potential  among solutions to the equations of motion of the heterotic supergravity. One will be able to determine the true vacuum in string theory by minimizing the potentials in the sectors of type I and II in the same way, and comparing the values of the potentials.

\section{Conclusion and Discussion}\label{sec:discussion}
\setcounter{equation}{0}
In this paper, in string geometry theory, we 
have identified heterotic perturbative vacua in superstring theory, which include all the heterotic superstring backgrounds. A non-trivial part of the vacua $\bar{\bm G}_{dd}$ is identified as follows. We expand the action in string geometry theory up to the second order of the scalar mode fluctuations of the metric, corresponding to perturvative superstrings, around the perturbative vacua \eqref{eq:sec3_backgrounds} and \eqref{eq:sec3_condition2}. First, we imposed a condition \eqref{eq:sec3_f_diffeq}, where the first order terms of the fluctuations vanish. This condition  means that 
$\bar{\bm G}_{dd}$, corresponding to the fluctuations is on-shell. 
 Second, we imposed an additional condition for $\bar{\bm G}_{dd}$ (\ref{eq:sec3_phi_equation}). Under this condition, we have derived the path-integrals of  pertrubative heterotic superstrings up to any order on the general heterotic superstring backgrounds from two-point correlation functions obtained from the second order terms of the  fluctuations.

  We have also obtained a potential for heterotic string backgrounds by substituting the heterotic string vacua \eqref{eq:sec3_backgrounds} and \eqref{eq:sec3_condition2} into the ``classical'' potential in string geometry theory and imposing the conditions \eqref{eq:sec3_f_diffeq} and  (\ref{eq:sec3_phi_equation})   by the method of Lagrange multipliers.  For applications for string phenomenology, we have further obtained a potential on a restricted region of string backgrounds with the warped compactification.
The authors in \cite{Nagasaki:2023fnz} conjecture that the potential for string backgrounds in the whole sector of string geometry theory represents the string theory landscape and can determine the true vacuum in string theory.

Here, we discuss the difference between the potential for heterotic string backgrounds obtained in this paper and the low-energy effective potentials in string theory. The action of string geometry theory (\ref{action of bos string-geometric model}) is a fundamental one that formulates string theory non-perturbatively. Thus, its potential terms  can determine a true vacuum in string theory. In this paper, we restrict the potential to heterotic perturbative vacua, which include heterotic string backgrounds, and call it a potential for heterotic string backgrounds.   A true heterotic vacuum will be determined by minimizing the potential among the heterotic string backgrounds that satisfy  consistency conditions of the string perturbations (Weyl invariance).   
On the other hand, 
we can obtain the low-energy effective action (potential), namely heterotic supergravity just by interpreting the  consistency conditions as equations of motion of it. Thus, the low-energy effective potential cannot determine a true vacuum in string theory by its minimum.  Actually, we impose the consistency conditions by solving them or by using the method of Lagrange multiplier to the potential for heterotic string backgrounds.

Next step is to search for the global minimum of the potential in string geometry theory.
That is, we will determine an internal geometry and fluxes. 
Among solutions to the equations of motion of the heterotic supergravity, the minimum of the potential (\ref{eq:sec5_diffeq_wcomp_f_1}) is one of the candidates of  the true vacuum in string theory. The true vacuum will be determined by searching for the minimum of the potentials for  type I and type II string backgrounds \cite{Kudo} in a similar way and comparing the values of the potentials. We will be able to search for the true vacuum in string theory without assuming naturalness or anthropic principle.  
One of the best analytic methods is to assume a class of internal spaces, especially Calabi-Yau manifolds and flux compactifications \cite{Masuda}, and then find the minimum in such a restricted region.    As a first step, the authors in \cite{Takeuchi} study a region of simple heterotic string phenomenological models and show that the minimum of the potential in this region has consistent phenomenological properties. This fact supports that the conjecture in \cite{Nagasaki:2023fnz} is correct. One of the best general methods is to  discretize the potential by the Regge calculus, and then find the minimum numerically \cite{Tanaka}. The fluctuations around the determined true vacuum will give the Standard Model in the four dimensions plus its corrections and an inflation in the early Universe.

\section*{Acknowledgements}
We would like to thank 
H. Kawai,
T. Masuda,
Y. Sugimoto,
M. Takeuchi,
G. Tanaka, 
K. Uzawa,
T. Yoneya,
and especially 
A. Tsuchiya
for discussions.

\appendix

\section{Derivations in detail}
\setcounter{equation}{0}
In this appendix, we will derive  in detail some formulas which we skipped in the main text. 

First, we will derive (\ref{eq:sec3_DeltaF_1}) from (\ref{eq:sec3_def_DeltaF}). 
By inserting 
\begin{subequations}
\begin{align}
1&= \int d\bar{\bm E}_m d\bm X_{G,m}(\bar\tau_m)d\lambda_{G,m}(\bar\tau_m) d\bar\tau_m
\big|\bar{\bm E}_m, \bm X_{G,m}(\bar\tau_m),\lambda_{G,m}(\bar\tau_m), \bar\tau_m \big\rangle\nonumber\\
&\hspace{15ex}
e^{-i\tilde\psi^\dagger_m\cdot\tilde\psi_m - \tilde\lambda_{Gm}\cdot\tilde\lambda_{G,m}}
\big<\bar{\bm E}_m, \bm X_{G,m}(\bar\tau_m),\lambda_{G,m}(\bar\tau_m), \bar\tau_m\big|, \\ 
1 &= \int  dp_{X_m} \left| p_{X_m}\right>\left<p_{X_m}\right|,\nonumber\\
1 &= \int dp_{\bar\tau_m}|p_{\bar\tau_m}\rangle\langle p_{\bar\tau_m}|,\nonumber\\
1 &= \int dp_{T_m}|p_{T_m}\rangle\langle p_{T_m}|,
\end{align}
\end{subequations}
\begin{align}
&\Delta_F(\bm X_{G,f}; \bm X_{G,i},\lambda_{G,f};\lambda_{G,i}|\bm E_f ; \bm E_i) \nonumber\\
&\coloneqq \int _0^\infty dT\big<\bm X_{G,f}, \lambda_{G,f}\:|\:\bm E_f; \bm E_i\big\Vert_{\rm out}  
 e^{-T\hat H}\big\Vert\bm X_{G,i}\lambda_{G,i} \,|\,\bm E_f ; \bm E_i\big>_{\rm in}\nonumber\\
&= \int _0^\infty dT\lim_{N\to\infty}\int_{\bm E_i}^{\bm E_f}\mathcal D\bm E\int_{\bm E_i}^{\bm E_f} \mathcal D\bm E' 
\prod_{n=1}^N \int d \bar{\bm E}_{n} d\bm X_{Gn}(\bar\tau_n)d\lambda_{G,n}(\bar\tau_n) d\bar\tau_n
e^{-i\tilde{\psi}^\dagger_n\cdot\tilde{\psi}_n - i\tilde{\lambda}_{G,n}^\dagger\cdot\tilde{\lambda}_{G,n}}
\nonumber \\
& \quad
 \prod_{m=0}^N\big<\bar{\bm E}_{m+1},\bm X_{G,m+1}(\bar\tau_{m+1}),\lambda_{G,m+1}(\bar\tau_{m+1}), \bar\tau_{m+1} \big| 
 e^{-\frac1N T\hat H}\big|\bar{\bm E}_m, \bm X_{G,m}(\bar\tau_m),\lambda_{G,m}(\bar\tau_{m}), \bar\tau_{m} \big> \nonumber \\
&= \int _0^\infty dT_0 \lim_{N\to\infty} \int dT_{N+1}
\int_{\bm E_i}^{\bm E_f}\mathcal D\bm E  
\prod_{m=1}^N \prod_{i=0}^N
\int dT_m d\bm X_{G,m}(\bar\tau_m)d\lambda_{G,m}(\bar\tau_m) d\bar\tau_m
 e^{-i\tilde\psi^\dagger_m \cdot \tilde{\psi}_m - i\tilde{\lambda}_{G,m}^\dagger\cdot\tilde{\lambda}_{G,m}}
\nonumber \\
&\hspace{5ex}
\int dp_{X}^i dp_{\bar{\tau}}^i
\big<X_{i+1}, \bar{\tau}_{i+1} \big\vert p_{X}^i p_{\bar{\tau}}^i \big>\big<p_{X}^i , p_{\bar{\tau}}^i \big\vert\big<\psi_{i+1}, \lambda_{G,i+1}\big\vert
e^{-\frac{1}{N}T_i \hat H}
\vert\psi_i,\lambda_{G,i}\big>\big\vert X_i, \bar{\tau}_i \big> \delta(T_i - T_{i+1})  \nonumber \\
&= \int _0^\infty dT_0 \lim_{N\to\infty} \int dT_{N+1}\int_{\bm E_i}^{\bm E_f}\mathcal D\bm E
\prod_{m=1}^N \prod_{i=0}^N
\int dT_m  d\bm X_{G,m}(\bar\tau_m)d\lambda_{G,m}(\bar\tau_m)  d\bar{\tau}_m
 e^{-i\tilde\psi^\dagger_m\cdot\tilde\psi_m - i\tilde{\lambda}_{G,m}^\dagger\cdot\tilde{\lambda}_{G,m}}
\nonumber \\
&\hspace{5ex}
\int dp_{X}^i  dp_{\bar{\tau}}^i e^{-\frac1N T_i H( p_{\bar{\tau}}^i, p_X^i, \sqrt{\bar h}G_{\mu\nu}(X_i(\bar{\tau}_i))\,\psi_i^\nu(\bar\tau_i), \bm X_{G,i}(\bar\tau_i),\lambda_{G,i}(\bar\tau_i), \bar{\bm E})}
e^{i\tilde\psi_{i+1}^\dagger\cdot\tilde\psi_i - i\tilde{\lambda}_{G, i+1}^\dagger\cdot\tilde{\lambda}_{G, i}}
\delta(T_i - T_{i+1})  \nonumber \\
&\hspace{5ex}
e^{ip_{X}^i\cdot(X_{i+1}-X_{i})
+i p_{\bar{\tau}}^i (\bar{\tau}_{i+1}-\bar{\tau}_i)}
\nonumber \\
&= \int _0^\infty dT_0 \lim_{N\to\infty} d T_{N+1} \int_{\bm E_i}^{\bm E_f} \mathcal D\bm E
 \prod_{n=1}^N \int d T_n  d\bm X_{G,n}(\bar\tau_n)d\lambda_{G,n}(\bar\tau_n) d\bar\tau_n
 \prod_{m=0}^N \int dp_{T_m}  dp_{X_m}(\bar\tau_m) dp_{\bar\tau_m} \nonumber\\
&\quad
 \exp\bigg[\sum_{m=0}^N \Delta t
\Big(-ip_{T_m} \frac{T_m - T_{m+1}}{\Delta t} 
- i\int d\bar{\sigma} \bar e p_{X_m}(\bar\tau_m)\frac{X_m(\bar\tau_m)-X_{m+1}(\bar\tau_{m+1})}{\Delta t} 
-ip_{\bar{\tau}_m} \frac{\bar{\tau}_m - \bar{\tau}_{m+1}}{\Delta t} 
\nonumber \\
&\qquad
 + i\tilde{\psi}^\dagger_{m+1} \frac{\tilde{\psi}_m(\bar\tau_m) - \tilde{\psi}_{m+1}(\bar\tau_{m+1})}{\Delta t}
 + i\tilde{\lambda}_{G,m+1}^\dagger \frac{\tilde{\lambda}_{G,m}(\bar\tau_m) - \tilde{\lambda}_{G,m+1}(\bar\tau_{m+1})}{\Delta t}\nonumber\\
&\qquad
 + T_m H\big(p_{\bar{\tau}_m}, p_{X_m}(\bar\tau_m),  \sqrt{\bar h}G_{\mu\nu}(X_{m}(\bar\tau_m))\,\psi^\nu_m(\bar\tau_m), \bm X_{G,m}(\bar\tau_m),\lambda_{G,m}(\bar\tau_m),\bar{\bm E}\big)\Big)\bigg]\nonumber\\
&= \int_{\bm E_i \bm X_{G,i},\lambda_{G,i}}^{\bm E_f, \bm X_{G,f},\lambda_{G,f}} 
 \mathcal D\bm E\mathcal D\bm X_{G}(\bar\tau)\mathcal D\lambda_{G}(\bar\tau)
  \mathcal{D}\bar{\tau}  \int \mathcal DT\int\mathcal Dp_T\mathcal Dp_X(\bar\tau)\mathcal{D}p_{\bar{\tau}}\nonumber\\
&\quad 
\exp\bigg[\int_{-\infty}^\infty dt\Big(
 -ip_T(t) \frac{d}{dt}T(t) 
 -i p_{\bar{\tau}}(t)\frac{d}{dt}\bar{\tau}(t)
 - i\int d\bar\sigma\bar e p_{X\mu}(\bar\tau(t), t) \frac{d}{dt} X^{\mu}(\bar\tau(t), t)\nonumber\\
&\qquad 
 + \int d\bar\sigma\frac{i}{2}\sqrt{\bar h}\Big(G_{\mu\nu}(X(\bar\tau(t), t))\,\bar\psi^\mu(\bar\tau(t), t)
\bar E^0_z\frac{d}{dt}\psi^\nu(\bar\tau(t), t)
 + \lambda^A_{G}(\bar\tau(t),t)\bar E^0_z\frac{d}{dt}\lambda_{G,A}(\bar\tau(t),t)\Big)
\nonumber \\
&\qquad 
 + T(t) H\big((p_{\bar{\tau}}(t), p_X(\bar\tau(t), t), \sqrt{\bar h}G_{\mu\nu}(X(\bar\tau(t), t))\,\psi^\nu(\bar\tau(t), t), \bm X_{G}(\bar\tau(t), t), \bar{\bm E}\big)\Big)\bigg],  
\end{align}
where  $\bar{\bm E}_0 = \bar{\bm E}'$, $\bm X_{G,0}(\bar\tau_0)= \bm X_{G,i}$, 
$\lambda_{G,0}(\bar\tau_0) = \lambda_{G,i}$,
$\bar\tau_0 = -\infty$, $\bar{\bm E}_{N+1} = \bar{\bm E}$,
$\bm X_{G,N+1}(\bar\tau_{N+1}) = \bm X_{G,f}$, 
$\lambda_{G,N+1}(\bar\tau_{N+1}) = \lambda_{G,f}$, $\bar\tau_{N+1} = \infty$, and 
$\Delta t\coloneqq 1/\sqrt{N}$.
A trajectory of points 
$[\bar{\bm\Sigma}, \bm X_{\hat D_{T}}(\bar\tau),\lambda_{G}(\bar\tau), \bar\tau]$ is necessarily continuous in $\mathcal M_D$ so that the kernel 
\begin{equation}
\big<\bar{\bm E}_{m+1}, \bm X_{G,m+1}(\bar\tau_{m+1}),\lambda_{G,m+1}(\bar\tau_{m+1}), \bar\tau_{m+1} \big| e^{-\frac1NT \hat H} \big|\bar{\bm E}_m,\bm X_{G,m}(\bar{\tau}_m),\lambda_{G,m}(\bar{\tau}_m), \bar{\tau}_m \big>
\end{equation}
in the fourth line is non-zero when $N \to \infty$.

Next, we will show that $t$ can be fixed to $\bar{\tau}$ by using a reparametrization of $t$ that parametrizes a trajectory in (\ref{eq:sec3_DeltaF_action}) with (\ref{eq:sec3_Lt}) and obtain (\ref{eq:sec3_DeltaF_result}) with (\ref{eq:sec3_action_result}).
In (\ref{eq:sec3_DeltaF_action}) with (\ref{eq:sec3_Lt}), the reparametrization invariance is fixed to a certain gauge. From now on, we will deform it to the theory without gauge fixing. After that, we will fix the reparametrization invariance to another  gauge, $ t=\bar{\tau}$. 
By inserting
$\displaystyle\int \mathcal Dc \mathcal Db\; e^{\int_0^1 dt(\frac{db(t)}{dt} \frac{dc(t)}{dt})},$
where $b(t)$ and $c(t)$ are $bc$-ghost, we obtain 
\begin{align}
&\Delta_F(\bm X_{G,f}; \bm X_{G,i}|\bm E_f ; \bm E_i)\nonumber\\
&= Z_0\int_{\bm E_i,\bm X_{G,i},\lambda_{G,i}}^{\bm E_f,\bm X_{G,f},\lambda_{G,f}} 
\mathcal DT\mathcal D\bm E\mathcal D\bm X
   \mathcal{D}\bar{\tau}
   \mathcal D\lambda_G
\mathcal Dp_T\mathcal Dc \mathcal Db\nonumber\\
&\quad
\exp\bigg[-\int_{-\infty}^{\infty} dt \Big(
 -ip_{T}(t) \frac{d}{dt} T(t)  
 +\frac{e^{2\phi}}{2}\frac1{T(t)}\Big(\frac{d\bar\tau(t)}{dt}\Big)^2
 +\frac{d b(t)}{dt} \frac{d (T(t) c(t))}{dt}\nonumber \\
&\quad\quad
+ \epsilon \Big( \int d\bar{\sigma} \sqrt{\bar{h}}G_{\mu\nu}(X(\bar{\tau}(t), t)) \Big( 
\frac12\bar{h}^{00}\frac{1}{T(t)}\partial_{t} X^{\mu}(\bar{\tau}(t), t)\partial_{t} X^{\nu}(\bar{\tau}(t), t) \nonumber \\
&\qquad\qquad
+\bar{h}^{01}\partial_{t} X^{\mu}(\bar{\tau}(t), t)\partial_{\bar{\sigma}} X^{\nu}(\bar{\tau}(t), t) 
+\frac12\bar{h}^{11}T(t)\partial_{\bar{\sigma}} X^{\mu}(\bar{\tau}(t), t)\partial_{\bar{\sigma}} X^{\nu}(\bar{\tau}(t), t)\Big)\nonumber\\
&\qquad
+ \int d\bar{\sigma}\,iB_{\mu\nu} (X(\bar{\tau}(t), t))
\partial_{t} X^{\mu}(\bar{\tau}(t), t)\partial_{\bar{\sigma}} X^{\nu}(\bar{\tau}(t), t)\nonumber\\
&\quad
 + \frac{i}{2}\int d\bar\sigma\sqrt{\bar h}G_{\mu\nu}\psi^\mu
 (\bar E^0_z\partial_t + T\bar E^1_z\partial_{\bar\sigma})\psi^\nu
\nonumber\\
&\quad
 + \frac{i}{2}\int d\bar\sigma\sqrt{\bar h}
 (\bar E^0_z\partial_t + T\bar E^1_z\partial_{\bar\sigma})X^\mu\psi^\nu\Big(\Gamma_{\nu\mu\rho} - \frac12H_{\nu\mu\rho}\Big)\psi^\rho
 + \frac12\int d\bar\sigma\sqrt{\bar h}
 (\bar E^0_z\partial_t + T\bar E^1_z\partial_{\bar\sigma})X^\mu\psi_\mu\chi_z\nonumber\\
&\quad
 + \frac{i}{2}\int d\bar\sigma\sqrt{\bar h}\lambda_G^A
 (\bar E^0_z\partial_t + T\bar E^1_z\partial_{\bar\sigma})\lambda_{G A}
 - \frac12\int d\bar\sigma\sqrt{\bar h}
 (\bar E^0_z\partial_t + T\bar E^1_z\partial_{\bar\sigma})X^\mu\lambda_G^AA_{\mu,AB}\lambda_G^B\nonumber\\
&\quad 
 - \frac{i}{4} T\int d\bar\sigma\sqrt{\bar h}F_{\mu\nu,AB}\psi^\mu\psi^\nu\lambda_G^A\lambda_G^B
 + \int d\bar\sigma\sqrt{\bar h}\frac{\gamma}{2\pi}TR_{\bar h}\Big)\Big)\bigg], 
\label{PropWMult}
\end{align}
where we redefine as $c(t) \to T(t) c(t)$, and $Z_0$ represents an overall constant factor. In the following, we will rename it $Z_1, Z_2, \cdots$ when the factor changes. The integrand variable $p_T (t)$ plays the role of the Lagrange multiplier providing the following condition,
\begin{align}\label{F1gauge}
F_1(t)\coloneqq \frac{d}{dt}T(t) = 0,
\end{align}
which can be understood as a gauge fixing condition. Indeed, by choosing this gauge in
\begin{align}
&\Delta_F(\bm X_{G,f}; \bm X_{G,i}|\bm E_f; \bm E_i) \nonumber\\
&= Z_1 \int_{\bm E_i,\bm X_{G,i},\lambda_{G,i}}^{\bm E_f,\bm X_{G,f},\lambda_{G,f}} 
\mathcal DT\mathcal D\bm E\mathcal D\bm X
\mathcal D\bar{\tau}
\mathcal D\lambda_G\nonumber\\
&\qquad
\exp\bigg[-\int_{-\infty}^{\infty} dt \Big(
 \frac{e^{2\phi}}{2}\frac1{T(t)}\Big(\frac{d\bar\tau(t)}{dt}\Big)^2
+\epsilon\Big(\int d\bar{\sigma} \sqrt{\bar{h}} G_{\mu\nu}(X(\bar{\tau}(t), t)) 
\Big(\frac12\bar h^{00}\frac{1}{T(t)}\partial_{t} X^{\mu}(\bar{\tau}(t), t)\partial_{t} X^{\nu}(\bar{\tau}(t), t)\nonumber \\
&\qquad\qquad
+ \bar{h}^{01}\partial_{t} X^{\mu}(\bar{\tau}(t), t)\partial_{\bar{\sigma}} X^{\nu}(\bar{\tau}(t), t) 
+ \frac12\bar{h}^{11}T(t)\partial_{\bar{\sigma}} X^{\mu}(\bar{\tau}(t), t)\partial_{\bar{\sigma}} X^{\nu}(\bar{\tau}(t), t)
\Big)\nonumber \\
&\qquad
 + \int d\bar{\sigma}iB_{\mu\nu} (X(\bar{\tau}(t), t))
\partial_{t} X^{\mu}(\bar{\tau}(t), t)\partial_{\bar{\sigma}} X^{\nu}(\bar{\tau}(t), t)\nonumber \\
&\quad
 + \frac{i}{2}\int d\bar\sigma\sqrt{\bar h}G_{\mu\nu}\psi^\mu
 (\bar E^0_z\partial_t + T\bar E^1_z\partial_{\bar\sigma})\psi^\nu
\nonumber\\
&\quad
 + \frac{i}{2}\int d\bar\sigma\sqrt{\bar h}
 (\bar E^0_z\partial_t + T\bar E^1_z\partial_{\bar\sigma})X^\mu\psi^\nu\Big(\Gamma_{\nu\mu\rho} - \frac12H_{\nu\mu\rho}\Big)\psi^\rho
 + \frac{1}{2}\int d\bar\sigma\sqrt{\bar h}
 (\bar E^0_z\partial_t + T\bar E^1_z\partial_{\bar\sigma})X^\mu\psi_\mu\chi_z\nonumber\\
&\quad
 + \frac{i}{2}\int d\bar\sigma\sqrt{\bar h}\lambda_G^A
 (\bar E^0_z\partial_t + T\bar E^1_z\partial_{\bar\sigma})\lambda_{G A}
 - \frac12\int d\bar\sigma\sqrt{\bar h}
 (\bar E^0_z\partial_t + T\bar E^1_z\partial_{\bar\sigma})X^\mu\lambda_G^AA_{\mu,AB}\lambda_G^B\nonumber\\
&\quad 
 - \frac{i}{4} T\int d\bar\sigma\sqrt{\bar h}F_{\mu\nu,AB}\psi^\mu\psi^\nu\lambda_G^A\lambda_G^B
 + \int d\bar\sigma\sqrt{\bar h}\frac{\gamma}{2\pi}TR_{\bar h}\Big)\Big)\bigg], 
\label{pathint2}
\end{align}
we obtain \eqref{PropWMult}.
The expression \eqref{pathint2} has a manifest one-dimensional diffeomorphism symmetry with respect to $t$, where $T(t)$ is transformed as an einbein \cite{Schwinger0}.

Under $d\bar\tau/d\bar\tau' = T(t)$, which implies 
\begin{equation}
\bar h^{00} = T^2\bar h'^{00},\quad
\bar h^{01} = T\bar h'^{01},\quad
\bar h^{11} = \bar h'^{11},\quad
\sqrt{\bar h} = \frac1T\sqrt{\bar h'}, \quad
\bar E^0_z = T \bar E'^0_z,
\end{equation}
we obtain 
\begin{align}
&\Delta_F(\bm{X}_{\hat{D}_{G}f}; \bm{X}_{\hat{D}_{G}i}|\bm{E}_f ; \bm{E}_i) \nonumber \\
&= Z_2 \int_{\bm E_i,\bm X_{G,i},\lambda_{G,i}}^{\bm E_f,\bm X_{G,f},\lambda_{G,f}} 
\mathcal DT\mathcal D\bm E\mathcal D\bm X
\mathcal D\bar{\tau}
\mathcal D\lambda_G
\nonumber \\
&\quad
\exp\bigg[-\int_{-\infty}^{\infty} dt \Big(
T(t)\frac{e^{2\phi}}{2}\Big(\frac{d\bar\tau(t)}{dt}\Big)^2
+\epsilon\Big(\int d\bar{\sigma} \sqrt{\bar h}G_{\mu\nu}(X(\bar{\tau}(t), t))
\Big( 
\frac12\bar{h}^{00}\partial_tX^{\mu}(\bar{\tau}(t), t)\partial_tX^{\nu}(\bar{\tau}(t), t)\nonumber\\ 
&\qquad\qquad
+ \bar{h}^{01}\partial_tX^{\mu}(\bar\tau(t), t)\partial_{\bar{\sigma}} X^{\nu}(\bar{\tau}(t), t) 
+ \frac{1}{2}\bar{h}^{11}\partial_{\bar{\sigma}} X^{\mu}(\bar{\tau}(t), t)\partial_{\bar{\sigma}} X^{\nu}(\bar{\tau}(t), t)
\Big)\nonumber \\
&\quad
+ \int d\bar{\sigma}\,iB_{\mu\nu} (X(\bar{\tau}(t), t))
\partial_{t} X^{\mu}(\bar{\tau}(t), t)\partial_{\bar{\sigma}} X^{\nu}(\bar{\tau}(t), t)\nonumber \\
&\quad
 + \frac{i}{2}\int d\bar\sigma\sqrt{\bar h}G_{\mu\nu}\psi^\mu
 (\bar E^0_z\partial_t + \bar E^1_z\partial_{\bar\sigma})\psi^\nu
\nonumber\\
&\quad
 + \frac{i}{2}\int d\bar\sigma\sqrt{\bar h}
 (\bar E^0_z\partial_t + \bar E^1_z\partial_{\bar\sigma})X^\mu\psi^\nu\Big(\Gamma_{\nu\mu\rho} - \frac12H_{\nu\mu\rho}\Big)\psi^\rho
 + \frac{1}{2}\int d\bar\sigma\sqrt{\bar h}
 (\bar E^0_z\partial_t + \bar E^1_z\partial_{\bar\sigma})X^\mu\psi_\mu\chi_z\nonumber\\
&\quad
 + \frac{i}{2}\int d\bar\sigma\sqrt{\bar h}\lambda_G^A
 (\bar E^0_z\partial_t + \bar E^1_z\partial_{\bar\sigma})\lambda_{G A}
 - \frac12\int d\bar\sigma\sqrt{\bar h}
 (\bar E^0_z\partial_t + \bar E^1_z\partial_{\bar\sigma})X^\mu\lambda_G^AA_{\mu,AB}\lambda_G^B\nonumber\\
&\quad 
 - \frac{i}{4} \int d\bar\sigma\sqrt{\bar h}F_{\mu\nu,AB}\psi^\mu\psi^\nu\lambda_G^A\lambda_G^B
 + \int d\bar\sigma\sqrt{\bar h}\frac{\gamma}{2\pi}R_{\bar h}\Big)\Big)\bigg],
\label{pathint3}
\end{align}
where $T(t)$ disappears except in front of the $\Big(\frac{d\bar\tau(t)}{dt}\Big)^2$ term.
This action is still invariant under the diffeomorphism with respect to t if $\bar{\tau}$ transforms in the same way as $t$. 

\begin{figure}[h]
\centering
\includegraphics[width=6cm]{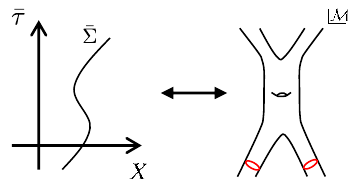}
\caption{A path and a  super Riemann surface. The line on the left is a trajectory in the path integral. The trajectory parametrized by $\bar{\tau}$ from $-\infty$ to $\infty$, represents a super Riemann surface with fixed punctures in $\mathcal M$ on the right.} 
\label{fig:pathintegral}
\end{figure}

If we choose a different gauge
\begin{equation}
F_2(t)\coloneqq \bar{\tau}(t)-t=0, \label{F2gauge}
\end{equation} 
in \eqref{pathint3}, we obtain 
\begin{align}
&\Delta_F(\bm X_{G,f}; \bm X_{G,i}| \bm E_f; \bm E_i)\nonumber\\
&= Z_3 \int_{\bm E_i,\bm X_{G,i},\lambda_{G,i}}^{\bm E_f,\bm X_{G,f},\lambda_{G,f}} 
\mathcal D\bm E\mathcal D\bm X\mathcal D\lambda_G
\mathcal D\alpha\mathcal Dc \mathcal Db\nonumber\\
&\quad
\exp\bigg[-\int_{-\infty}^{\infty} dt \Big(\alpha(t) (\bar{\tau}-t) +b(t)c(t)\Big(1-\frac{d \bar{\tau}(t)}{dt}\Big)+T(t)\frac{e^{2\phi}}{2}\nonumber\\
&\qquad
+\epsilon \Big(\int d\bar{\sigma} \sqrt{\bar{h}}G_{\mu\nu}(X(\bar{\tau}(t), t))\Big( 
\frac12\bar{h}^{00}\partial_{t} X^{\mu}(\bar{\tau}(t), t)\partial_{t} X^{\nu}(\bar{\tau}(t), t) 
\nonumber \\
&\qquad\qquad
+\bar{h}^{01}\partial_{t} X^\mu(\bar{\tau}(t), t)\partial_{\bar\sigma} X^\nu(\bar\tau(t), t) 
+\frac{1}{2}\bar{h}^{11}\partial_{\bar\sigma} X^\mu(\bar{\tau}(t), t)\partial_{\bar{\sigma}} X^{\nu}(\bar{\tau}(t), t)\Big)  \nonumber \\
&\qquad
+ \int d\bar{\sigma}iB_{\mu\nu} (X(\bar{\tau}(t), t))
\partial_t X^\mu(\bar{\tau}(t), t)\partial_{\bar{\sigma}} X^{\nu}(\bar{\tau}(t), t) \nonumber \\
&\quad
 + \frac{i}{2}\int d\bar\sigma\sqrt{\bar h}G_{\mu\nu}\psi^\mu
 (\bar E^0_z\partial_t + \bar E^1_z\partial_{\bar\sigma})\psi^\nu
\nonumber\\
&\quad
 + \frac{i}{2}\int d\bar\sigma\sqrt{\bar h}
 (\bar E^0_z\partial_t + \bar E^1_z\partial_{\bar\sigma})X^\mu\psi^\nu\Big(\Gamma_{\nu\mu\rho} - \frac12H_{\nu\mu\rho}\Big)\psi^\rho
 + \frac{1}{2}\int d\bar\sigma\sqrt{\bar h}
 (\bar E^0_z\partial_t + \bar E^1_z\partial_{\bar\sigma})X^\mu\psi_\mu\chi_z\nonumber\\
&\quad
 + \frac{i}{2}\int d\bar\sigma\sqrt{\bar h}\lambda_G^A
 (\bar E^0_z\partial_t + \bar E^1_z\partial_{\bar\sigma})\lambda_{G A}
 - \frac12\int d\bar\sigma\sqrt{\bar h}
 (\bar E^0_z\partial_t + \bar E^1_z\partial_{\bar\sigma})X^\mu\lambda_G^AA_{\mu,AB}\lambda_G^B\nonumber\\
&\quad 
 - \frac{i}{4} \int d\bar\sigma\sqrt{\bar h}F_{\mu\nu,AB}\psi^\mu\psi^\nu\lambda_G^A\lambda_G^B
 + \int d\bar\sigma\sqrt{\bar h}\frac{\gamma}{2\pi}R_{\bar h}\Big)\Big)\bigg].
\nonumber \\
&= \int_{\bm E_i,\bm X_{G,i},\lambda_{G,i}}^{\bm E_f,\bm X_{G,f},\lambda_{G,f}} 
\mathcal D\bm E \mathcal D\bm X\mathcal D\lambda_G
\nonumber \\
&\quad
\exp\bigg[-\epsilon\int_{-\infty}^{\infty} d\bar\tau\Big(
\int d\bar{\sigma} \sqrt{\bar{h}}G_{\mu\nu}(X(\bar{\tau}(t), t))  \Big( 
\frac12\bar{h}^{00}\partial_t X^{\mu}(\bar{\sigma}, \bar{\tau})\partial_{\bar{\tau}} X^{\nu}(\bar{\sigma}, \bar{\tau})
 \nonumber \\
&\qquad\qquad
 + \bar{h}^{01}\partial_t X^{\mu}(\bar{\sigma}, \bar{\tau})\partial_{\bar\sigma} X^{\nu}(\bar\sigma, \bar\tau) 
 + \frac12\bar{h}^{11}\partial_{\bar\sigma} X^{\mu}(\bar\sigma, \bar\tau)\partial_{\bar\sigma} X^{\nu}(\bar{\sigma}, \bar{\tau})\Big)
 \nonumber \\
&\qquad
+ \int d\bar{\sigma}iB_{\mu\nu} (X(\bar{\sigma}, \bar{\tau}))
\partial_t X^{\mu}(\bar{\sigma}, \bar{\tau})\partial_{\bar{\sigma}} X^{\nu}(\bar{\sigma}, \bar{\tau}) 
\nonumber \\
&\quad
 + \frac{i}{2}\int d\bar\sigma\sqrt{\bar h}G_{\mu\nu}\psi^\mu
 (\bar E^0_z\partial_t + \bar E^1_z\partial_{\bar\sigma})\psi^\nu
\nonumber\\
&\quad
 + \frac{i}{2}\int d\bar\sigma\sqrt{\bar h}
 (\bar E^0_z\partial_t + \bar E^1_z\partial_{\bar\sigma})X^\mu\psi^\nu\Big(\Gamma_{\nu\mu\rho} - \frac12H_{\nu\mu\rho}\Big)\psi^\rho
 + \frac{1}{2}\int d\bar\sigma\sqrt{\bar h}
 (\bar E^0_z\partial_t + \bar E^1_z\partial_{\bar\sigma})X^\mu\psi_\mu\chi_z\nonumber\\
&\quad
 + \frac{i}{2}\int d\bar\sigma\sqrt{\bar h}\lambda_G^A
 (\bar E^0_z\partial_t + \bar E^1_z\partial_{\bar\sigma})\lambda_{G A}
 - \frac12\int d\bar\sigma\sqrt{\bar h}
 (\bar E^0_z\partial_t + \bar E^1_z\partial_{\bar\sigma})X^\mu\lambda_G^AA_{\mu,AB}\lambda_G^B\nonumber\\
&\quad 
 - \frac{i}{4} \int d\bar\sigma\sqrt{\bar h}F_{\mu\nu,AB}\psi^\mu\psi^\nu\lambda_G^A\lambda_G^B
 + \int d\bar\sigma\sqrt{\bar h}\frac{\gamma}{2\pi}R_{\bar h}\Big)\bigg],
\label{prelastaction}
\end{align}
where  we have redefined as $T(t)\frac{e^{2\phi}}{2} \to T'(t)$ and integrated out $T'(t)$. 
The path integral is defined over all possible two-dimensional super Riemannian manifolds with fixed punctures in the manifold $\mathcal{M}$ defined by the metric $G_{\mu\nu}$, as in Fig.\ref{fig:pathintegral}.

The fields in (\ref{prelastaction}) are the representatives with respect to  the super  diffeomorphism times super Weyl invariance.
Because the action  in (\ref{prelastaction}) has those symmetries, 
the representatives can be transformed to the general fields: 
\begin{equation}
\Delta_F(\bm X_{G,f};\bm X_{G,i}| \bm E_f;\bm E_i)
= Z\int_{\bm E_i,\bm X_{G,i},\lambda_{G,i}}^{\bm E_f,\bm X_{G,f},\lambda_{G,f}}
\mathcal D\bm E\mathcal D\bm X\mathcal D\lambda_G e^{-\gamma \chi}e^{-S_{\rm s}}, 
\end{equation}
where 
\begin{align}
S_{\rm s} &= \frac1{2\pi\alpha'}\int d\sigma\sqrt{h(\sigma,\tau)}\Big(
 \big(h^{mn}(\sigma,\tau)G_{\mu\nu}(X(\sigma,\tau)) 
  + i\varepsilon^{mn}(\sigma,\tau)B_{\mu\nu}(X(\sigma,\tau))\big)\partial_m X^\mu\partial_n X^\nu\nonumber\\
&\quad
 + iE^a_zG_{\mu\nu}(X(\sigma,\tau))\psi^\mu(\sigma,\tau)\partial_a\psi^\nu(\sigma,\tau)\nonumber\\
&\quad
 + iE^a_z\partial_aX^\mu(\sigma,\tau)\psi^\nu(\sigma,\tau)
 \Big(\Gamma_{\nu\mu\rho} - \frac12H_{\nu\mu\rho}(X(\sigma,\tau))\Big)\psi^\rho(\sigma,\tau)
 + E^a_z\partial_a X^\mu(\sigma,\tau)\psi_\mu(\sigma,\tau)\chi_z(\sigma,\tau)\nonumber\\
&\quad
 + iE^a_z\lambda_G^A\big(\partial_a\lambda_{G A} (\sigma,\tau) 
  +i \partial_aX^\nu(\sigma,\tau) A_{\nu,AB}\lambda_G^B(\sigma,\tau)\big)
 - \frac{i}{2}F_{\mu\nu,AB}\psi^\mu(\sigma,\tau)\psi^\nu(\sigma,\tau)
  \lambda_G^A(\sigma,\tau)\lambda_G^B(\sigma,\tau)\Big), 
\end{align}
and $\chi$ is the Euler number of the two-dimensional Riemannian manifold. In order to set our scale the string scale, we have deleted $\epsilon$ and introduced $\alpha'$ in front of the action, by rescaling the fields of the target coordinates $X^\mu$, $\lambda_G^A$ and $\psi^\mu$.
For regularization,  by renormalizing $\bm\psi_{dd}''$, we divide the correlation function by the constant factor $Z$ and by the volume of the super diffeomorphism times the super Weyl transformation 
$V_{{\rm diff}\times{\rm Weyl}}$.

\end{document}